\documentclass[journal=jpccck,manuscript=article]{achemso}

\usepackage{chemformula} 
\usepackage[T1]{fontenc} 
\usepackage{tabularx}
\usepackage{chemmacros}
\usepackage{booktabs}
\usepackage{siunitx}
\usepackage{multirow}
\usepackage{graphicx}
\usepackage{subcaption}
\usepackage{soul}
\usepackage[version=4]{mhchem}
\usepackage{lineno} 
\usepackage{array}
\usepackage{changes}
\usepackage[export]{adjustbox} 

\newcolumntype{Y}{>{\centering\arraybackslash} m{4.5cm} >{\centering\arraybackslash} m{1.5cm}}

\newcolumntype{X}{>{\centering\arraybackslash} m{4.5cm} >{\centering\arraybackslash} m{1.5cm}} 

\newcolumntype{W}{>{\centering\arraybackslash} m{3cm} >{\centering\arraybackslash} m{5mm}}
\author{Stephan Mohr}
\affiliation{Nextmol (Bytelab Solutions SL), Barcelona, Spain}
\alsoaffiliation{Barcelona Supercomputing Center (BSC)}
\email{stephan.mohr@nextmol.com}

\author{Felix Hoevelmann}
\affiliation{Clariant Produkte (Deutschland) GmbH, Frankfurt, Germany}

\author{Jonathan Wylde}
\affiliation{Clariant Oil Services, Clariant Corporation, Houston, Texas, USA}
\alsoaffiliation{Heriot Watt University, Edinburgh, Scotland, UK}

\author{Natascha Schelero}
\affiliation{Clariant Produkte (Deutschland) GmbH, Frankfurt, Germany}

\author{Juan Sarria}
\affiliation{Clariant Produkte (Deutschland) GmbH, Frankfurt, Germany}

\author{Nirupam Purkayastha}
\affiliation{Clariant Produkte (Deutschland) GmbH, Frankfurt, Germany}

\author{Zachary Ward}
\affiliation{Clariant Oil Services, Clariant Corporation, Houston, Texas, USA}

\author{Pablo Navarro Acero}
\affiliation{Nextmol (Bytelab Solutions SL), Barcelona, Spain}

\author{Vasileios K. Michalis}
\affiliation{Barcelona Supercomputing Center (BSC)}

\title{Ranking the Efficiency of Gas Hydrate Anti-agglomerants Through Molecular Dynamic Simulations}

\abbreviations{IR,NMR,UV}

\begin{document}

\begin{tocentry}
\includegraphics[width=1.0\textwidth]{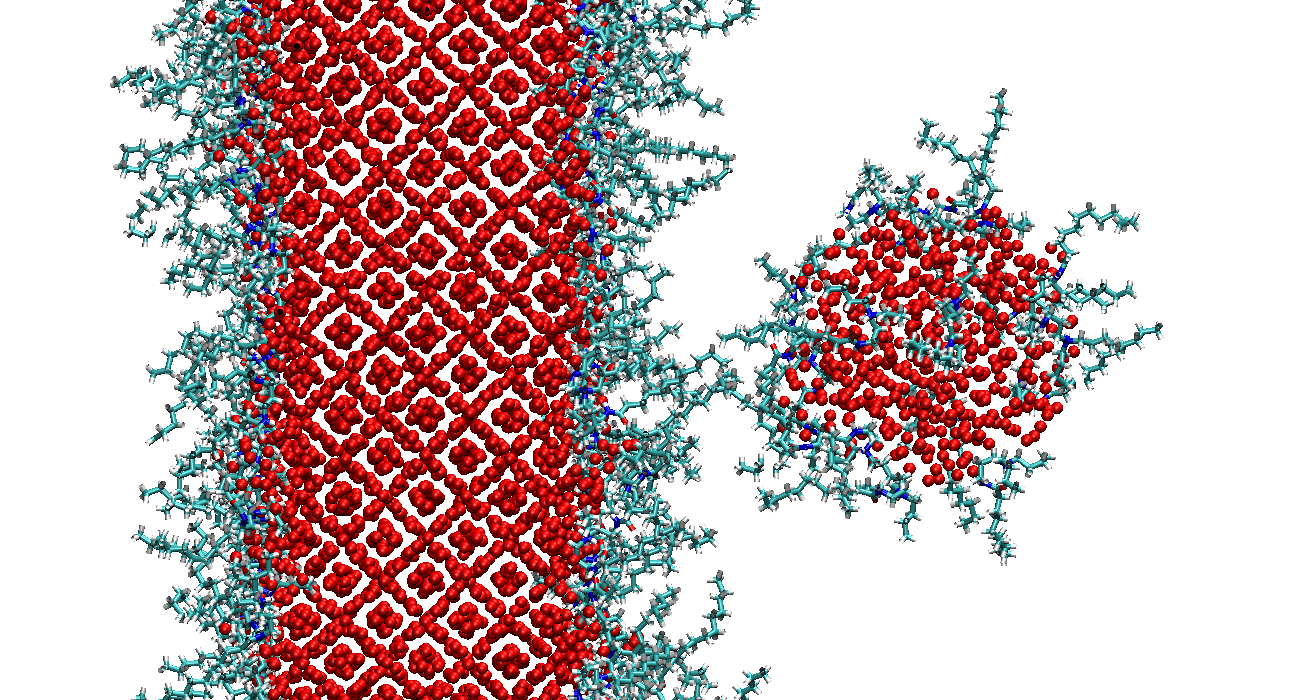}
\end{tocentry}

\begin{abstract}
 Using both computational and experimental methods, the capacity of four different surfactant molecules to inhibit the agglomeration of sII hydrate particles was assessed. The computational simulations were carried out using both steered and non-steered Molecular Dynamics (MD), simulating the coalescence process of a hydrate slab and a water droplet, both covered with surfactant molecules. The surfactants were ranked according to free energy calculations (steered MD) and number of agglomeration events (non-steered MD). The experimental work was based on rocking cell measurements, determining the minimum effective dose necessary to inhibit agglomeration. Overall, good agreement was obtained between the performance predicted by the simulations and the experimental measurements. Moreover, the simulations allowed to gain additional insights that are not directly accessible via experiments, such as an analysis of the mass density profiles, the diffusion coefficients, or the orientations of the long tails.
\end{abstract}

\section{Introduction}
Gas hydrates~\cite{Sloan2007-Clathrate} or clathrates are inclusion compounds that consist of water and low molecular weight molecules, typically hydrocarbon species,  being enclathrated in crystalline three dimensional cages that the water molecules form around them via hydrogen bonds. The encaged molecules are customarily called guests and more than 130 different species can form gas hydrates, such as methane, ethane, propane, nitrogen, carbon dioxide, and hydrogen. Depending on the size of the guest as well as its interactions with the water, different crystal structures can be formed, most common being the sI, sII and sH, which differ in the type and ratio of cages. For example, the unit cell of sII hydrates consists of 16 dodecahedron cages (5\textsuperscript{12}) and 8 hexakaidecahedron cages (5\textsuperscript{12}6\textsuperscript{4}).

The natural abundance of gas hydrates~\cite{Boswell2011-Current} and their industrial importance primarily in the oil and gas industry~\cite{Zerpa2011-Surface} has motivated gas hydrate research for many decades~\cite{Koh2002-Towards}. Of particular importance is the research direction on hydrate inhibition~\cite{Kelland2006-History}, since formation of gas hydrates during oil and gas production operations poses as serious flow assurance concern as they can spontaneously form under common operational conditions causing pipeline and equipment blockages that can yield catastrophic effects.
The formation of gas hydrates is a complex multiscale problem, ranging from molecular interactions at the sub-nanometer scale over the formation of the first particles (micrometer scale), their agglomeration (millimeter scale) up to the effects on the flowline (meter scale). An extensive discussion on the topic can be found in a recent series of papers by Bassani \latin{et al.}~\cite{Bassani2019-A_Multiscale_1, Bassani2020-A_Multiscale_2, Bassani2020-A_Multiscale_3}.
Apart from reducing the water cut whenever possible, there are two main hydrate prevention methods. The first method is via the so-called thermodynamic inhibitors, such as methanol or ethylene glycol, which shift the phase coexistence line of the system outside the operational pressure and temperature boundaries. The second method relies on the addition of a variety of molecules, at low dosage rates~\cite{Kelland2006-History, Perrin2013-The_chemistry}, that can either delay hydrate nucleation (kinetic inhibitors) or prevent the agglomeration of formed hydrate particles for a period greater than the residence time inside the pipelines (anti-agglomerants, AAs).  Compared to kinetic inhibitors, AAs have the advantage of being applicable in a larger subcooling range, and are therefore better suited for deepwater applications~\cite{Guo2013-Offshore}.

There exist a number of commercial AA products that can effectively prevent the formation of hydrates in oil-water-gas mixtures~\cite{Kolotova2020-Evaluation}.
However, their action mechanism is not yet fully understood~\cite{Kelland2018-A_review}.
There is an increasing need to improve our understanding of the relevant systems, to produce more efficient and greener AAs and to drive the discovery of new suitable molecules through innovative approaches that go beyond the traditional trial and error based experimental methods. Atomistic simulation methods such as molecular dynamics (MD) and Monte Carlo (MC)~\cite{Frenkel2001-Understanding} can play such an innovative role as they can provide critical insights at a molecular level into the behavior of a wide range of fluids and materials. Both MD and MC methods have been used in numerous hydrate related studies~\cite{English2015-Perspectives,Barnes2013-Advances}. Nevertheless, only a limited number of these studies deals explicitly with low dosage hydrate inhibitors. Additionally, most of them are focused on kinetic inhibitors~\cite{Carver1995-Inhibition, Carver1996-Characterisation, Kvamme1997_Molecular, Freer2000-An_Engineering, Carver2000-Configuration, Storr2004-Kinetic, Kvamme2005-Molecular, Anderson2005_Properties, Hawtin2006_Polydispersity, Moon2007_Nucleation, Gomez2007_2007, Kuznetsova2010_Impact, Davenport2011_A_simple, Kuznetsova2012_Molecular, Yagasaki2015_Adsorption, Lim2016_Dual, Yagasaki2018_Adsorption, Bertolazzo2018_The_Clathrate, Yagasaki2019-Molecular}, whereas the number of atomistic simulation studies on hydrate AAs is even smaller~\cite{Phan2016-Molecular, Bui2017-Evidence, Sicard2018-Emergent, Bellucci2018-Molecular, JimenezAngeles2018-Hydrophobic, Mehrabian2018-Effect, Mehrabian2019-In_silico, Naulage2019-How, Bui2020-Synergistic}. However, all the studies on AAs have been published within the last few years, showing that hydrate anti agglomeration research based on atomistic methodologies is gaining momentum.

Phan \latin{et al.}~\cite{Phan2016-Molecular} employed steered and equilibrium MD simulations to investigate the coalescence of an sI hydrate particle and a water droplet embedded in a hydrocarbon mixture (n-decane and methane). They employed various inhibitors within the general category of quaternary ammonium chlorides and investigated whether their presence can prevent coalescence. In particular, they studied the effect of the head group (e.g. methyl versus butyl groups) and of the hydrophobic tail lengths (e.g. n-hexadecyl versus n-dodecyl tails). They concluded that when the water droplet is not covered by surfactants it is more likely to coalesce with the hydrate than when surfactants are present on both surfaces. Further they concluded that surfactants with butyl tripods on the quaternary head group and hydrophobic tails with size similar to the solvent molecules can act as effective anti-agglomerants by forming a protective film on the hydrate surface.

Bui \latin{et al.}~\cite{Bui2017-Evidence} used MD simulations to study the structure of thin films of AAs adsorbed at the interface between sII methane hydrate and a liquid hydrocarbon, composed of dissolved methane and higher-molecular-weight alkanes such as n-hexane, n-octane, and n-dodecane. The anti-agglomerants considered were ammonium chlorides with two long tails (either with 8 or 12 carbon atoms) and one short tail (either with four, six or eight carbon atoms). At low surface densities, the hydrophobic tails did not show a preferred orientation, irrespectively of the tail length. At sufficiently high surface densities, the simulations showed differences in the structure of the interfacial film depending on the features of the surfactant and on the type of hydrocarbons present in the system. Some anti-agglomerants were found to pack densely at the interface excluding methane from the interfacial region. The hydrophobic tails of the antiagglomerants that showed this feature have a length comparable to that of the n-dodecane in the liquid phase. The simulation results were compared against experimental data, and it was found that the anti-agglomerants that produced well-ordered films performed better in experiments. The methane exclusion effect of the AA films was further investigated by Sicard \latin{et al.}~\cite{Sicard2018-Emergent}.

Bellucci \latin{et al.}~\cite{Bellucci2018-Molecular} examined via MD the surface adsorption of a single n-dodecyl-tri(n-butyl)ammonium chloride on a sII methane-propane hydrate surface in contact with either an aqueous or a liquid hydrocarbon phase. They identified the preferred binding sites and binding configurations and measured through thermodynamic integration the relevant binding free energies. They noted differences in the binding mechanism between the configurations of the aqueous and hydrocarbon system and concluded that the inhibitor is less effective in the aqueous phase because the surface adsorption is less favorable. This leads to the assumption that AAs stabilize water-in-oil emulsions and destabilizes capillary liquid bridges between hydrate particles. Additionally, the authors concluded that the ammonium cation does not incorporate into the local water structure of the hydrate lattice upon binding due to steric hindrance stemming from the short hydrocarbon chains of the AA.

Using MD, Jim\'enez-\'Angeles and Firoozabadi~\cite{JimenezAngeles2018-Hydrophobic} studied the adsorption behavior of a single molecule on an aqueous-hydrate interface in the presence of NaCl at various concentrations. The molecules examined were n-decane, a nonionic surfactant (cocamidopropyl dimethylamine), a cationic surfactant (didodecyl dimethylammonium chloride), and an anionic surfactant (sodium dodecyl sulfate). The authors distinguished the binding of the various molecules: either via hydrophobic hydration where water molecules form a hydrogen bond network similar to clathrate hydrates, or via ionic hydration where water molecules align according to the polarity of an ionic group. Their analysis consisted of hydrogen bond and tetrahedral density profiles as a function of the distance to the chemical groups as well as potential of mean force profiles through steered simulations. They found that the non-ionic surfactant and the hydrocarbon chain induce hydrophobic hydration and are favorably adsorbed on the hydrate surface, whereas for the ionic surfactants adsorption is not favorable through the head. The addition of NaCl disrupts hydrophobic hydration, reduces the solubility of solutes in the aqueous solution and enhances the adsorption of the surfactants on the hydrate surface.

Mehrabian \latin{et al.}~\cite{Mehrabian2018-Effect} extended the previous work by Belluci \latin{et al.}~\cite{Bellucci2018-Molecular} by examining the surface adsorption of a n-dodecyl-tri(n-butyl)-ammonium chloride on an sII methane-propane hydrate in the case of different salinities. By analyzing the binding configurations and calculating the binding free energies, they concluded that the salt decreases the solubility of the AA, increases the thermodynamic driving force for adsorption, and additionally the salt ions create a negatively charged interfacial layer close to the hydrate surface that enhances the solvation of the cationic head of the AA. They also noted that the lowest free energy of binding occurs for the simultaneous head and tail binding configuration compared with the head-only or tail-only binding configuration. Mehrabian \latin{et al.}~\cite{Mehrabian2019-In_silico} also considered, for a system containing a 3.5 wt \% NaCl brine, the effect of the length of the long tail (ranging from eight to 16 carbon atoms) on the binding affinity through Umbrella Sampling-based potential of mean force profiles, and binding statistics based on brute force simulations. Interestingly, the authors note that these are extremely important in the sense that they can capture phenomena that sometimes might be missed from free energy calculations due to the necessary restrictions applied to the examined configuration space. The authors showed that the dodecyl tail provides the optimal balance between the enthalpic and entropic effects and has the highest binding affinity.

Continuing this direction of optimizing the long tails of n-dodecyl-tri(n-butyl)-ammonium chloride, Mehrabian \latin{et al.}~\cite{Mehrabian2020-In_Silico} used bias-exchange metadynamics simulations to sample the bindings configurations of AAs at the hydrate surface and to calculate their binding free energies.
In agreement with previous studies, they found that configurations where both the head and the tail of the AA are bound are the most favorable ones.
Furthermore, they engineered new AAs by replacing the dodecyl tail with more rigid structures (biphenyl, pyrene and fluorene), with the goal of decreasing the entropic penalty upon binding (and in turn decreasing the free energy of binding).
However, this modification of the rigidity decreased the binding free energy only little.
On the other hand, replacing a methylene group of the middle ring of fluorene with an oxygen atom lead to a considerabe decrease.

Naullage \latin{et al.}~\cite{Naulage2019-How} investigated the structure and dynamics of surfactant films at the hydrate-oil interface, and their impact on the contact angle and coalescence between hydrate particles  and water droplets. They found that surfactant-covered hydrate-oil interfaces are super-hydrophobic, but that a large contact angle is not sufficient to predict good anti agglomeration performance of a surfactant. They concluded that the length of the surfactant molecules, the density of the interfacial film, and the strength of binding of its molecules to the hydrate surface are the main factors in preventing coalescence and agglomeration of hydrate particles with water droplets.

Bui \latin{et al.}~\cite{Bui2020-Synergistic} investigated the synergistic and antagonistic effects that aromatic compounds, dissolved in the hydrocarbon phase, can have on the performance of anti-agglomerants. They concluded that polycyclic aromatics could enhance the performance of the specific surfactants that were considered, whereas monocyclic aromatics could, in some cases, negatively affect performance.

These atomistic studies provide unprecedented insights on the fundamental mechanisms of hydrate agglomeration and how AAs can prevent it.
On the other hand, this atomic resolution comes at a price, and such studies are currently limited to system sizes of the order of ten nanometers and simulation times of some hundred nanoseconds.
A possibility to increase both the length and time scales is to resort to coarse-grained simulations, such as the ones conducted by Molinero and co-workers~\cite{Jacobson2010-A_Methane-Water, Jacobson2010-Amorphous, Jacobson2011-Can, Bertolazzo2018_The_Clathrate, Naulage2019-How} using the monoatomic water model mW~\cite{Molinero2009-Water}.
Using such an approach can be a good compromise between the resolution required to capture the important molecular mechanisms and the need to perform large scale simulations in order to mimic realistic systems.

The aforementioned studies treat various different setups, for instance concerning the type of the hydrate (sI~\cite{Phan2016-Molecular, JimenezAngeles2018-Hydrophobic} or sII~\cite{Bui2017-Evidence, Sicard2018-Emergent, Bellucci2018-Molecular, Mehrabian2018-Effect, Mehrabian2019-In_silico, Mehrabian2020-In_Silico, Bui2020-Synergistic}), the occupation (pure methane~\cite{Phan2016-Molecular, Bui2017-Evidence, Sicard2018-Emergent, JimenezAngeles2018-Hydrophobic, Bui2020-Synergistic} or mixed methane-propane~\cite{Bellucci2018-Molecular, Mehrabian2018-Effect, Mehrabian2019-In_silico, Mehrabian2020-In_Silico}), or the interface (hydrate-hydrocarbon~\cite{Phan2016-Molecular, Bui2017-Evidence, Sicard2018-Emergent, Bellucci2018-Molecular, Bui2020-Synergistic} or hydrate-water~\cite{Bellucci2018-Molecular, JimenezAngeles2018-Hydrophobic, Mehrabian2018-Effect, Mehrabian2019-In_silico, Mehrabian2020-In_Silico}).
Overall, however, they clearly reflect two distinct approaches in the investigation of hydrate anti agglomeration, the first being the binding affinity of isolated AA molecules and the second the collective behavior of these molecules. The distinction between these approaches is dictated by the restrictions imposed by the currently available computational power and is expected to be lifted in the future. Although both approaches are equally important, the present study focuses on the second direction, namely that of the collective behavior of AA molecules preassembled on the hydrate surface. To this purpose we have decided to study a number of different AA molecules in a system that consists of a flat hydrate surface and a water droplet in contact with a liquid hydrocarbon phase, given that in hydrate anti agglomeration the oil phase serves as the medium where anti agglomeration takes place. It would appear as advantageous to study the behavior of hydrate particles suspended in a liquid hydrocarbon phase, but a considerable number of preliminary studies that we have carried out indicated that these hydrate particles are unstable at the conditions of interest (more details are given below) and have the tendency to melt below a radius of approximately \SI{6}{nm}. Given that systems containing hydrate particles with radii larger than \SI{6}{nm} are computationally out of reach for the present study, we have decided to work with a flat hydrate surface that interacts with a water nanodroplet, effectively mimicking the behavior of a large hydrate particle due to the liquid layer that is formed on the latter. The issue of instability of small hydrate nanoparticles can be attributed to the Gibbs-Thomson effect~\cite{Kaptay2012-The_Gibbs}, which describes the apparent shift of the melting temperature to lower values for particles that have high specific surface. This issue, although being important, is poorly examined in the literature, particularly for systems as the one used in the present study, and we are interested in devoting a separate study to this phenomenon. Alternatively, it would also be possible to use rigid hydrate particles, but this approach would prevent us from capturing the quasi-liquid water layer that can be formed on the hydrate surface and which can play an important role in the behavior of the AAs. This phenomenon, known as pre-melting, is well established for the case of ice~\cite{Conde2008-The_thickness, Slater2019-Surface} but poorly studied for the case of gas hydrates~\cite{Maeda2015-Is}, except for the sI methane hydrate - methane system~\cite{JimenezAngeles2014-Induced, Ding2008-Molecular}. In any case, thanks to our specific selection of the system setup where all molecules can move freely, this layer is included in the present study. 

The composition of the liquid hydrocarbon phase is also an issue that must be addressed. Given our goal to create a system as realistic as possible, we have selected a liquid hydrocarbon phase that is the product of the equilibrium between n-dodecane and green canyon gas, given that the green canyon gas is customarily employed in experimental hydrate inhibition setups. The green canyon gas contains multiple molecular species (see Table~\ref{tab:green_canyon_gas}), but the ones of greatest relevance are methane, ethane and propane. A model of similar composition has thus been implemented and allowed to equilibrate in contact with n-dodecane. The composition of the equilibrated liquid phase has then been used in the subsequent simulations (details are given below).

We have examined 4 different AA molecular species with clear differences in their chemistries, increasing the significance of the current study. More importantly the experimental behavior of these molecules is known, which allows us to rank them regarding their anti agglomeration potency, and thus to compare this experimentally based ranking to the computationally based ranking that we offer in this study. The purpose of this work is to provide a comparison between these two rankings and to explore thus the capability of computational methods to be applied to the prediction of the anti agglomeration behavior of different molecules. To this end we have carried out extensive MD simulations, both steered and non-steered, with multiple independent runs, paying attention to the intrinsically stochastic behavior of this kind of systems, and we have tried to gain insight into the apparently different behavior of the AAs through a number of relevant analyses.

The paper is organized in the following manner. Initially, details of the methodology are given, including the description and preparation of the structures, the employed force fields, the simulations and the analysis methods. Following, the main findings of this work are presented and discussed. Finally, the conclusions are summarized.

\section{Methodology}

\subsection{Computational setup}

\begin{figure}
 \includegraphics[width=1.0\textwidth]{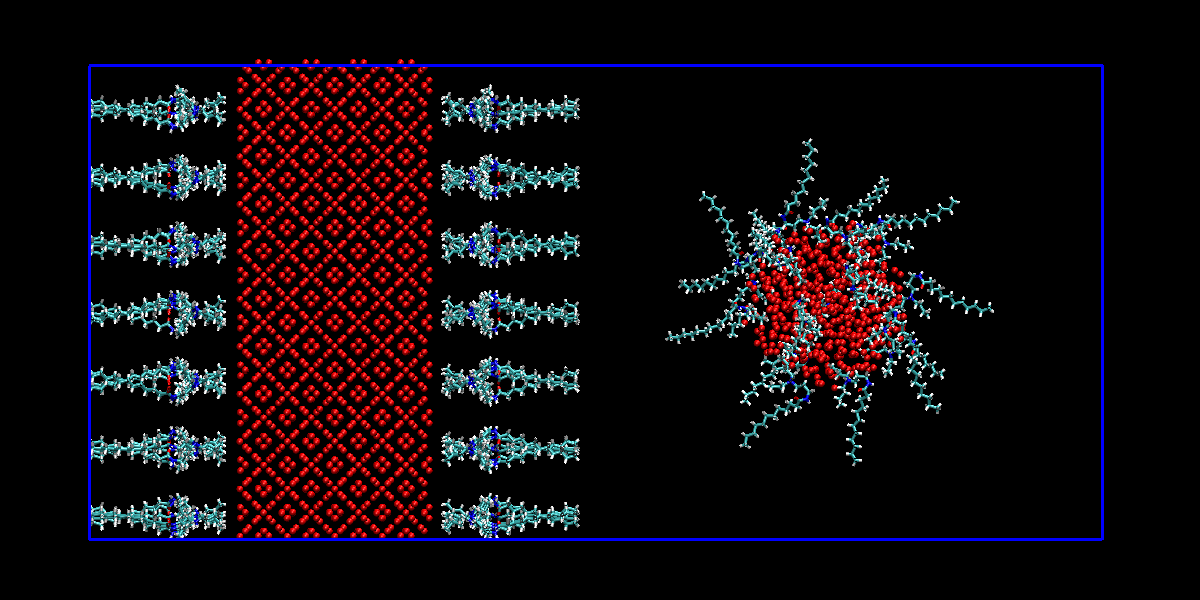}
 \includegraphics[width=1.0\textwidth]{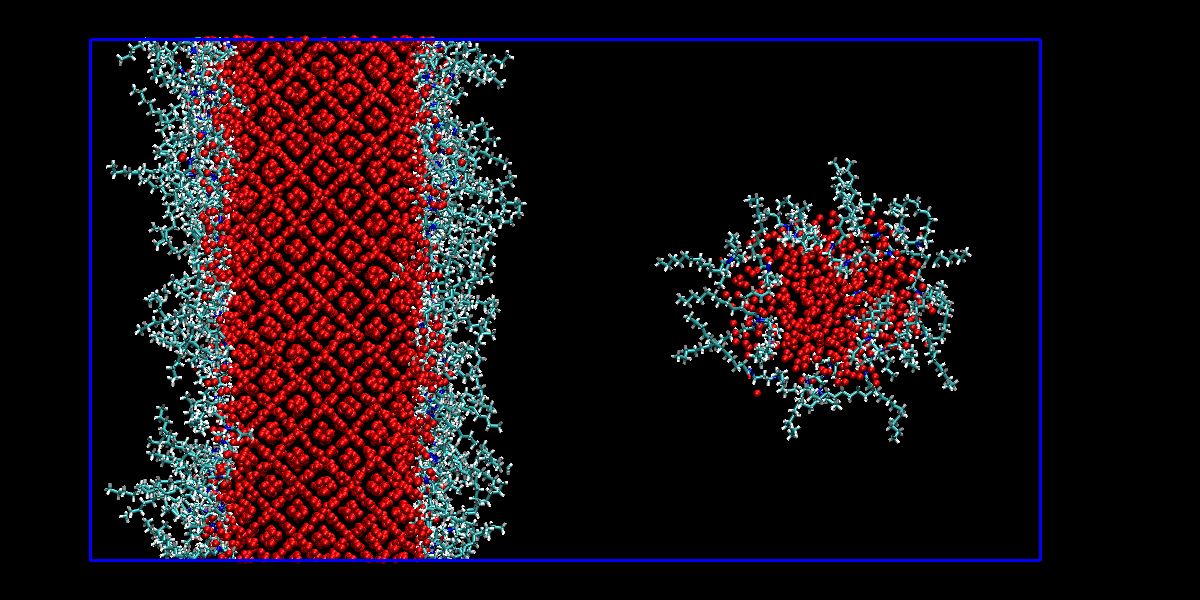}
 \caption{Two snapshots of the initial (top) and equilibrated (bottom) configurations of the system where both the hydrate and water droplet surfaces are covered with AA1 molecules at low concentration. The red spheres represent the oxygens of the water molecules, while a balls-and-sticks representation is used for the AA molecules. The molecules of the hydrocarbon phase are hidden for reasons of clarity.}
 \label{fig:system}
\end{figure}

The rectangular simulation system that we have implemented is composed of three components. At the bottom we placed a flat sII methane-propane hydrate slab in the form of a 5x5x2 supercell.  The positions of the hydrate atoms were taken from Takeuchi et al.~\cite{Takeuchi2013-Water}. All small cages were filled with methane molecules and all large cages with propane molecules. The hydrate slab thus consisted of 6800 water, 800 methane and 400 propane molecules. Subsequently, we covered the surface on both sides with the desired number of AAs, arranged in a regular grid and oriented perpendicular to the surface. Next, we created a water sphere with radius \SI{1.5}{nm} (505 molecules, carved out from a box of bulk water) and centered it at a distance of \SI{4.5}{nm} above the AA layer. Depending on the specific system, the water droplet was first covered with a layer of AAs, oriented perpendicular to the droplet surface. Finally, the entire system was solvated with a hydrocarbon mixture consisting of 50 mol\% dodecane (1700 molecules), 4 mol\% propane (136 molecules), 6 mol\% ethane (204 molecules) and 40 mol\% methane (1360 molecules). This molar composition corresponds to the equilibrium of pure dodecane in contact with simplified green canyon gas (see below). Typical box dimensions are \SI{8.6}{nm} in the x and y dimension, while the z dimension is of the order of \SI{16}{nm} to \SI{19}{nm}. Figure~\ref{fig:system} shows a snapshot of the initial configuration for a particular AA along with a snapshot of the same system after equilibration.

We examined four different AA molecules; their structures and IUPAC names are presented in Table~\ref{tab:molecules}. The first two ones, AA1 and AA2, are identical up to the counterion that is used. The third and fourth ones, AA3 and AA4, have a simpler structure (no spacer group), but contain three alkyl tails instead of two. Moreover, their counterion (chloride) is much smaller compared to AA1 and AA2. In fact, AA4 is a surfactant molecule that has not been designed as a hydrate anti-agglomerant, but we include it in our study due to its structural similarity (C12 long tail, ammonium head group with short tails) with the other AAs.

\begin{table}
 \begin{center}
 \begin{tabularx}{\textwidth}{m{0.12\textwidth} >{\centering}m{0.19\textwidth} >{\centering}m{0.19\textwidth} >{\centering}m{0.19\textwidth} m{0.19\textwidth}<{\centering}}
  \toprule  
  Structure
  &
  \includegraphics[height=0.22\textheight]{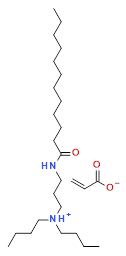}
  &
  \includegraphics[height=0.22\textheight]{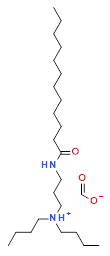}
  &
  \includegraphics[height=0.22\textheight]{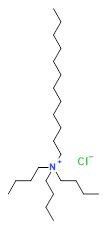}
  &
  \includegraphics[height=0.22\textheight]{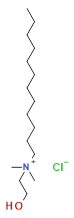}
  \\
  \midrule
  IUPAC name
  &
  \iupac{di|butyl-[3-(dodecanoyl|amino)propyl]ammonium acrylate}
  &
  \iupac{di|butyl-[3-(dodecanoyl|amino)propyl]ammonium formate}
  &
  \iupac{tri|butyl(dodecyl)ammonium chloride}
  &
  \iupac{dodecyl-(2-hydroxy|ethyl)-di|methyl-ammonium chloride}
  \\
  \midrule
  Short name
  &
  AA1
  &
  AA2
  &
  AA3
  &
  AA4
  \\
  \bottomrule
 \end{tabularx}
 \end{center}
 \caption{The four surfactants investigated in this paper.}
 \label{tab:molecules}
\end{table}

For each of the 4 different candidate AA molecules, we examined three different surface concentrations, with the water droplet either covered or not covered with the AA molecules. The hydrate surface was covered on each side with 49 molecules, 121 molecules and 196 molecules, for the low, medium and high concentration, respectively. These configurations correspond to surface coverages of 0.65, 1.62 and 2.62 molecules/nm\textsuperscript{2}. The molecules were placed on nodes of a rectangular grid (7x7, 11x11 and 14x14, respectively). For the covered water droplet cases, the AA surface concentration of the droplet was approximately equal to that of the hydrate surface: 18 (low), 46 (medium) and 74 (high) molecules. This corresponds to surface densities of 0.64, 1.63 and 2.62 molecules/nm\textsuperscript{2}, respectively. In total we therefore obtain 24 different systems.

The TIP4P/ice model~\cite{Abascal2005-A_potential} was used to represent water, whereas methane, ethane, propane and n-dodecane were described with TraPPE-UA~\cite{Potoff2001-Vapor}. The inhibitors were described using the GAFF force field~\cite{Wang2004-Development}. To obtain the GAFF parameters, each AA molecule was first optimized at the HF level (basis set 6-31G*) using the NWChem package (version 6.6)~\cite{Valiev2010-NWChem}, and then the topology was created using the Ambertools package (version 17)~\cite{Wang2006-Automatic}. The partial atomic charges were calculated using the AM1-BCC method~\cite{Jakalian2000-Fast, Jakalian2002-Fast}. All the cross-interaction parameters were calculated using the Lorentz-Berthelot combining rules~\cite{Frenkel2001-Understanding}.
This combination of force fields has already been used in other similar studies~\cite{Bui2017-Evidence, Sicard2018-Emergent, Bui2020-Synergistic}.

All simulations were carried out using the GROMACS MD simulation package (version 2018.3)~\cite{VanDerSpoel2005-GROMACS, Hess2008-GROMACS, Pronk2013-GROMACS, Abraham2015-GROMACS}. All the non-steered simulations were carried out in the isothermal-isobaric (NPT) ensemble, while the steered simulations were done in the canonical (NVT) ensemble. The Berendsen~\cite{Berendsen1984-Molecular} temperature and pressure coupling schemes were employed with time constants of \SI{0.5}{ps}. Anisotropic pressure coupling was used with equal compressibility in all directions, so that each dimension can fluctuate independently, to avoid inducing stresses to the hydrate crystal. The leap-frog integration algorithm was employed with a time step of \SI{2}{fs} and periodic boundary conditions were applied in all directions. The Lennard-Jones interactions were truncated at \SI{12}{\angstrom} without employing any dispersion corrections given that the system is anisotropic and inhomogeneous. The long-range Coulombic interactions were handled with the particle mesh Ewald (PME) method~\cite{Essmann1995-A_smooth}. In all simulations the pressure was set to \SI{100}{bar} and the temperature to \SI{277}{K}. For the steered simulations, we used a pulling velocity of \SI{0.05}{nm/ns} and a force constant of \SI{3000}{kJ.mol^{-1}.nm^{-2}} was applied to restrain the center of mass position of the water droplet along its path in the direction perpendicular to the surface.

Prior to the production runs, a three-step equilibration protocol was executed for all systems examined in the present study. Firstly, energy minimization was carried out with the steepest descent algorithm, keeping fixed the positions of the oxygen atoms and the methane and central propane atoms of the hydrate slab only. This initial step has the purpose of avoiding unfavorable overlaps during the initial placement of the molecules. Secondly, a \SI{50}{ps} NPT run with a time step of \SI{0.2}{fs} was applied to gently relax the box volume close to each final volume. Thirdly, a \SI{100}{ns} NVT run was carried out keeping again fixed the positions of the water oxygens and the methane and central propane atoms of the hydrate slab. This final step allows the AA molecules to relax on the hydrate surface. Subsequently all the production non-steered runs were carried out in the NPT ensemble without any constraints to any atomic position, whereas the steered runs were done in the NVT ensemble. The simulation time of the non-steered runs was \SI{600}{ns} (upon no coalescence of the water droplet with the hydrate surface occurred) except in the cases of multiple independent runs where the simulation time was \SI{400}{ns}. The simulation time of the steered runs was \SI{120}{ns}.

\begin{figure}
 \includegraphics[width=0.7\textwidth]{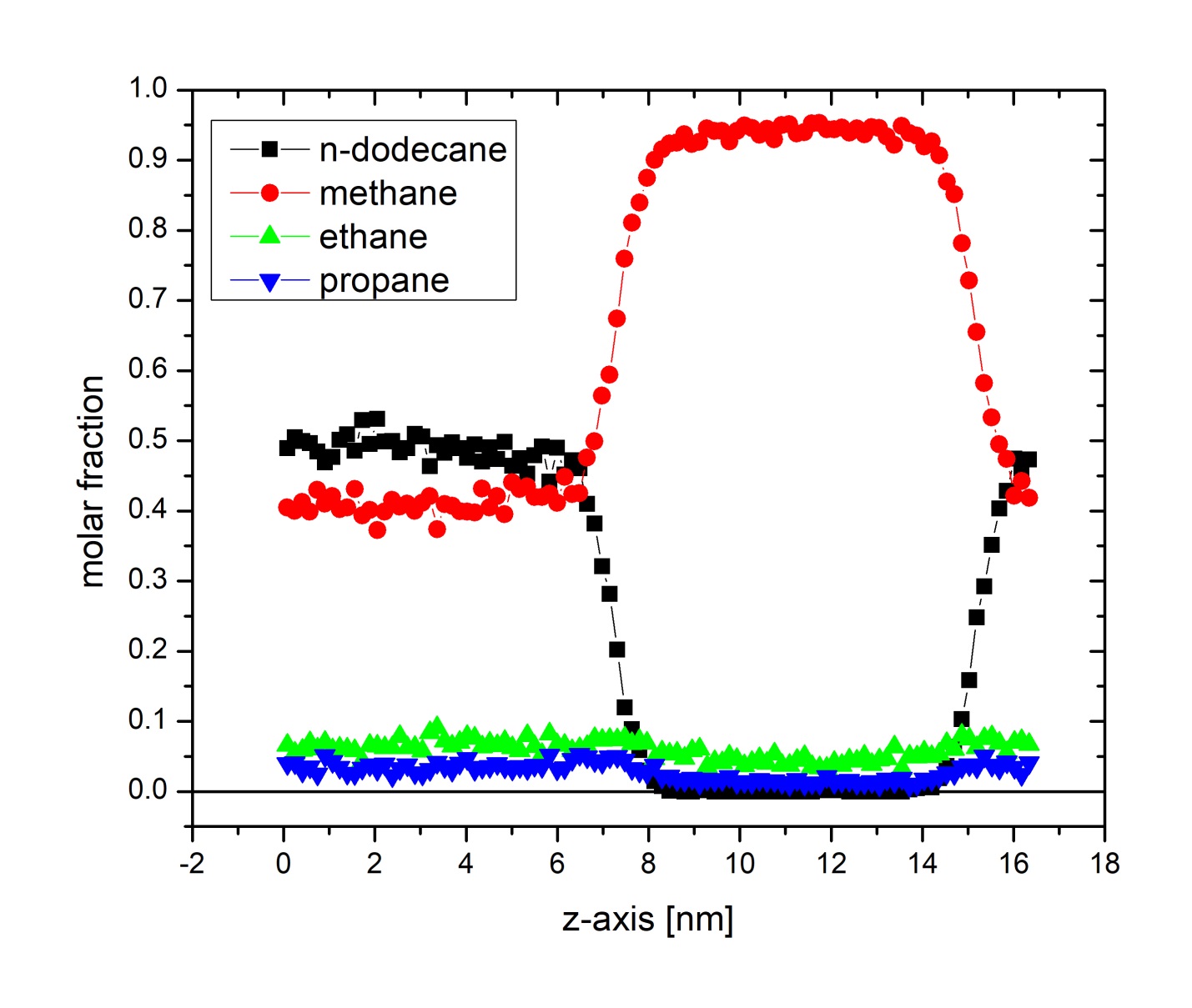}
 \caption{Molar fractions of the liquid and gaseous hydrocarbon phases that are the product of equilibration between a pure n-dodecane phase and a gaseous phase with an initial molar composition of 87.5\%, 8.375\% and 4.125\% in methane, ethane and propane, respectively. The boundary between the two phases is located at about \SI{7}{nm} and \SI{16}{nm} along the z dimension.}
 \label{fig:hydrocarbon_composition}
\end{figure}

The calculation of the composition of the hydrocarbon phase that was used in the present study was carried out via a direct coexistence simulation of a pure n-dodecane phase with a gaseous phase with a composition representative of green canyon gas (see Table~\ref{tab:green_canyon_gas}). The initial n-dodecane phase had 300 molecules, while the gaseous phase was composed of 700 methane, 67 ethane and 33 propane molecules. The duration of the run was \SI{200}{ns}, while pressure coupling was applied only to the direction normal to the interface, with the other two dimensions kept fixed. The equilibrated density profiles are presented in Figure~\ref{fig:hydrocarbon_composition}, with the resulting equilibrium composition of the liquid phase being 50 mol\% dodecane, 4 mol\% propane, 6 mol\% ethane and 40 mol\% methane.

\subsection{Experimental setup}
\label{sec:experimental_setup}

The 4 different anti-agglomerant molecules studied throughout this work were experimentally evaluated using a rocking cell test. The rocking cell test is a commonly used test for assessing the performance of anti-agglomerant chemistry. Briefly, additives are evaluated based on their ability to effectively minimize the size of hydrate particle agglomerates and then to disperse those particles into the hydrocarbon phase. The results were classified as ``pass'' or ``fail'' based on whether hydrate blockages were detected. Performance was evaluated by determining the minimum effective dose (MED) required to register as a ``pass'' in the rocking cell test. The effective dosages (MEDs) were screened for 5.0 wt\% NaCl brine at 50 vol.-\% watercut and \SI{138}{bar} at \SI{4}{\degreeCelsius}.

The rocking cell apparatus (``rack'') is comprised of a plurality of sapphire tubes, each placed within a stainless steel support cage. Each assembled sapphire tube and steel cage (hereby referred to as a rocking cell) is typically loaded with fluids containing a hydrocarbon fluid phase and a brine phase, along with a stainless steel ball for mixing. The rocking cell can withstand pressures of up to \SI{200}{bar} (\SI{2900}{psi}). The rocking cell, once loaded with the fluids, is then mounted on the rack with gas injection and pressure monitoring. During testing, as the gases cool and hydrates form, the consumed gas was replaced via a high-pressure syringe pump to maintain the system at constant pressure.

\begin{figure}
 \includegraphics[width=0.7\textwidth]{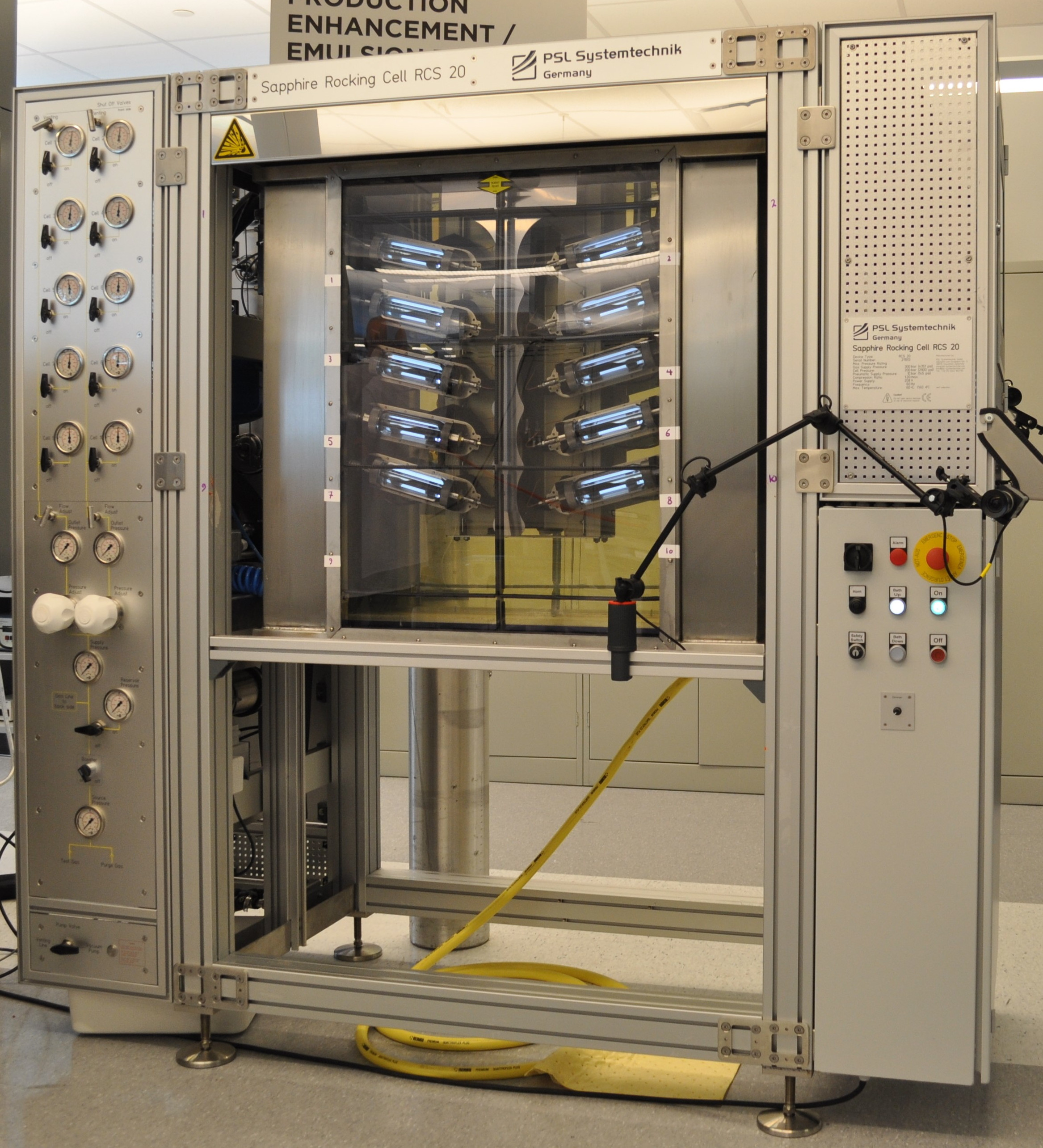}
 \caption{The rocking cell aparatus used for the experimental validation.}
 \label{fig:rocking_cell}
\end{figure}

The rack was loaded with two sets of rocking cells with an overall 2x5 sapphire tube configuration (two tubes wide and 5 tubes tall). The center position on the rack (between both cells) was fixed and allowed to rotate while the outer positions on the rack were moved vertically up and down. This vertical motion allowed the rocking cells to rotate into a positive or negative angle position. The steel ball placed inside the sapphire tube moved from one end of the cell to the other during a rocking motion. The rack rocked up and down at a rate of about 5 complete cycles (up and down) every minute. The rack was further contained within a temperature-controlled bath attached to a chiller with temperature control from \SI{-10}{\degreeCelsius} to \SI{60}{\degreeCelsius}. A picture of the apparatus is shown in Figure~\ref{fig:rocking_cell}.

The rocking cells were filled with three components: hydrocarbon, aqueous phase, and gas. First, each rocking sapphire tube was filled with \SI{5}{ml} of dodecane and a \SI{5}{ml} of 5\% NaCl brine (watercut 50 vol.-\%) for a total liquid loading of 50\% total tube volume (\SI{20}{ml} total). The inhibitor was added as a 50 wt.-\% active solution at dose rates in percent, by volume of water (vol.-\%). Green Canyon gas was used for this testing with its composition given in Table~\ref{tab:green_canyon_gas}. 

\begin{table}
 \begin{tabularx}{0.7\textwidth}{l l r}
  \toprule  
  Component Name & Chemical Symbol & Amount (mol-\%) \\
  \midrule
  Nitrogen       & \ce{N2}              & 0.14            \\
  Carbon Dioxide & \ce{CO2}             & 0.00            \\
  Methane        & \ce{C1}              & 87.56           \\
  Ethane         & \ce{C2}              & 7.60            \\
  Propane        & \ce{C3}              & 3.00            \\
  i-Butane       & \ce{i-C4}            & 0.50            \\
  n-Butane       & \ce{n-C4}            & 0.80            \\
  i-Pentane      & \ce{i-C5}            & 0.20            \\
  n-Pentane      & \ce{n-C5}            & 0.20            \\
  \bottomrule
 \end{tabularx}
 \caption{Green Canyon gas composition.}
 \label{tab:green_canyon_gas}
\end{table}

To determine the effectiveness of the various concentrations of anti-agglomerants, we executed the following rocking cell test procedure:

\begin{enumerate}
 \item Pretest Steps: Once the rack has been loaded with the rocking cells containing hydrocarbon fluid and brine, the rocking cells are evacuated with a vacuum pump for 15-20 minutes. While evacuating, the bath temperature is increased to the starting test temperature of 49\si{\degree}C. Once the bath has reached 49\si{\degree}C, the cells and the syringe pump are pressurized with Green Canyon gas to 138 bar and the syringe pump is switched on to maintain pressure during initial saturation.
 \item Saturation Step: The apparatus is set to rock at 5 rocks per minute for 2 hours to ensure the hydrocarbon fluids and brine loaded in the cell have been saturated with gas. This testing is performed at constant pressure and the syringe pump remains switched on and set at 138 bar for the remainder of the test.
 \item Cooling Step: While maintaining a rocking rate of 5 rocks per minute, the system is cooled from \SI{49}{\degree}C to \SI{4}{\degree}C over 6 hours.
 \item Steady State Mixing Step before Shut-in: At the constant temperature of \SI{4}{\degree}C, the apparatus is kept rocking at 5 rocks per minute for 12 hours to ensure complete hydrate formation.
 \item Shut-in Step: The apparatus is set to stop rocking and to set the cell position to horizontal and kept at a constant temperature of \SI{4}{\degree}C for 12 hours.
 \item Steady State Mixing Step after Shut-in: At the conclusion of the shut in period, the apparatus is restarted at the rate of 5 rocks per minute at the constant temperature of \SI{4}{\degree}C for 4 hours.
 \item Test Completion: At the conclusion of the experiment, the apparatus is set to stop rocking and the cells are set at a negative inclination to keep fluids away from the gas injection port. The chiller bath is set to \SI{49}{\degree}C to melt any formed hydrates and allow for depressurization and cleaning.
\end{enumerate}

\begin{figure}
 \begin{subfigure}{.49\textwidth}
  \includegraphics[width=1.0\linewidth]{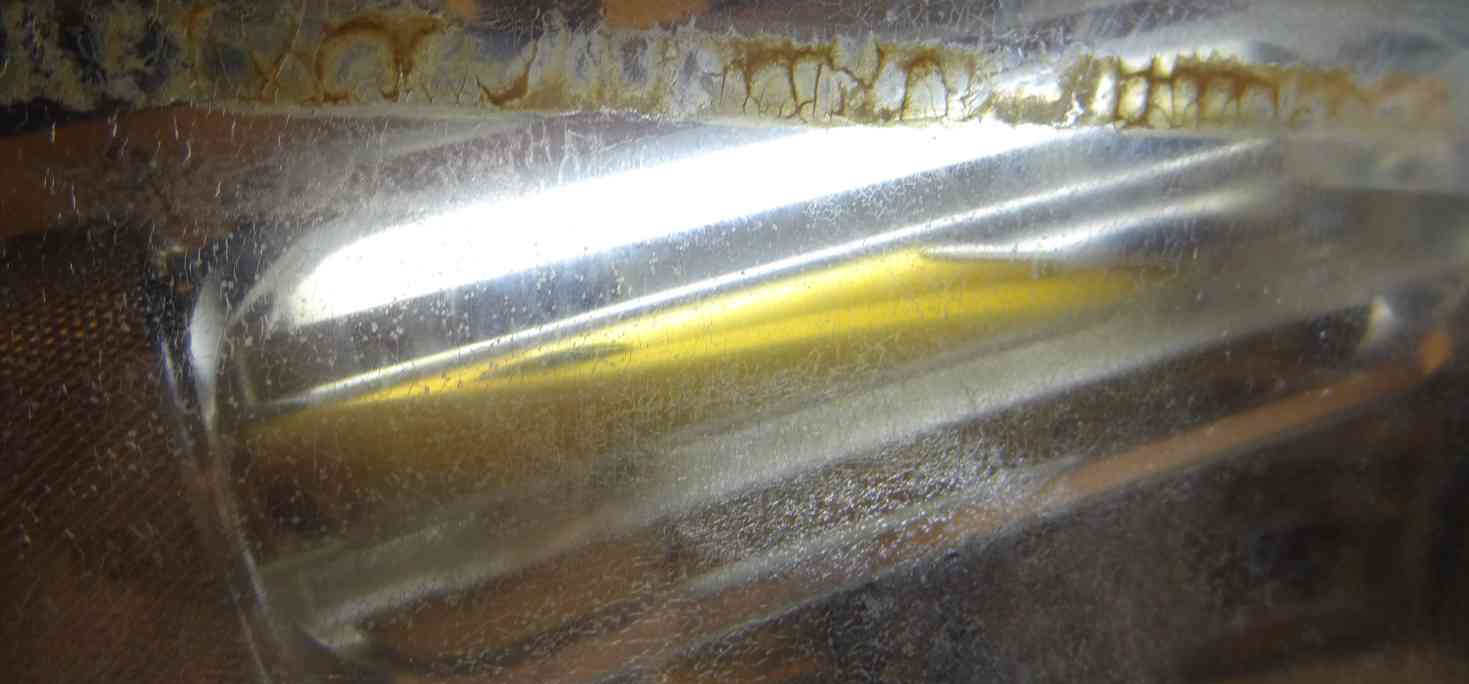}
  \caption{Well dispersed hydrate slurry.}
  \label{fig:picture_cell_dose20}
 \end{subfigure}
 \begin{subfigure}{.49\textwidth}
  \includegraphics[width=1.0\linewidth]{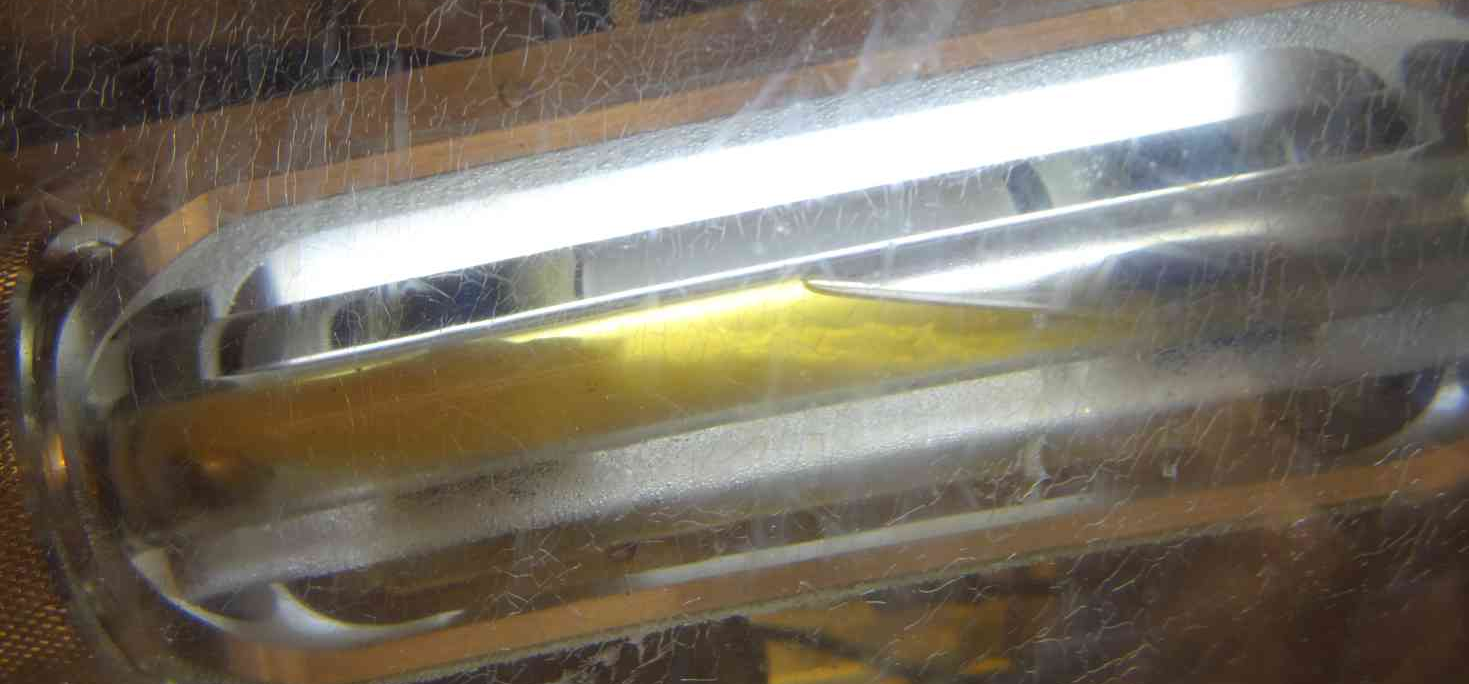}
  \caption{Hydrate slurry with some clumping.}
  \label{fig:picture_cell_dose15}
 \end{subfigure}
 \begin{subfigure}{.49\textwidth}
  \includegraphics[width=1.0\linewidth]{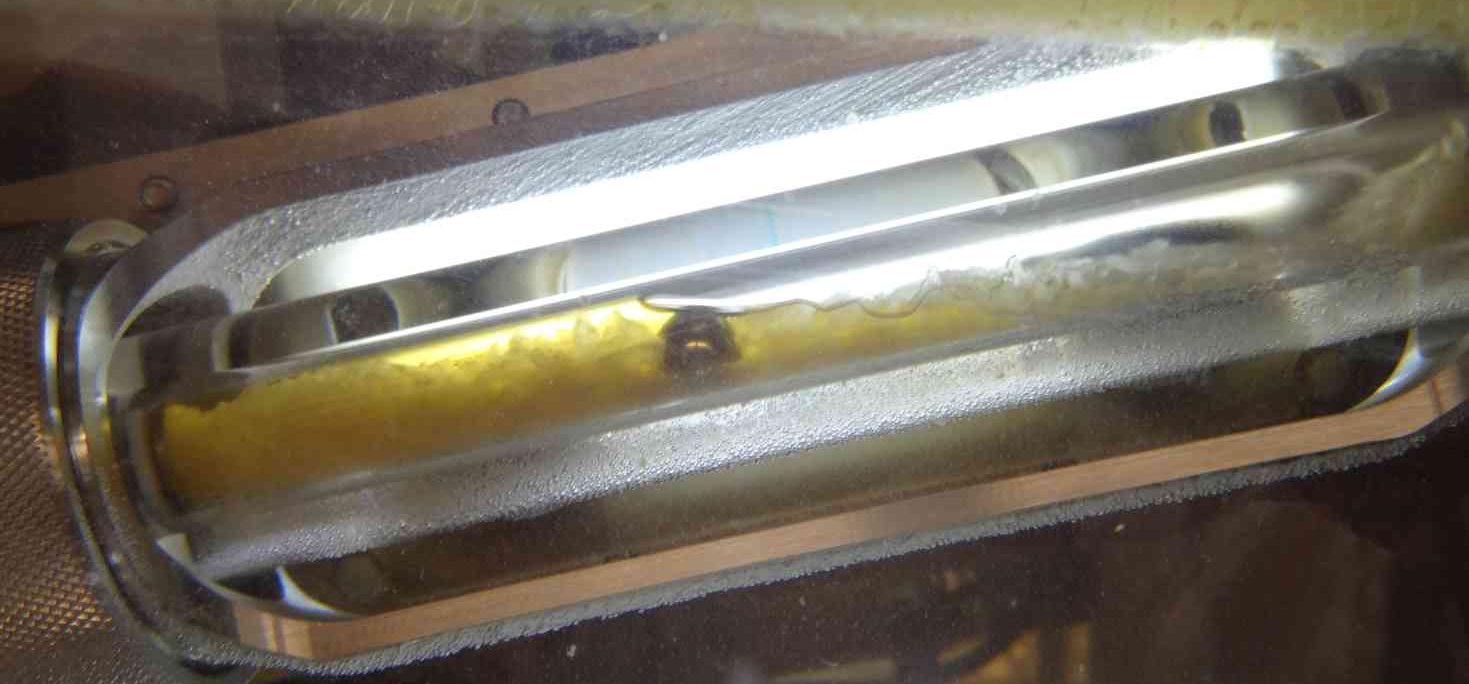}
  \caption{Clumping and some hydrate plugs.}
  \label{fig:picture_cell_dose10}
 \end{subfigure}
 \begin{subfigure}{.49\textwidth}
  \includegraphics[width=1.0\linewidth]{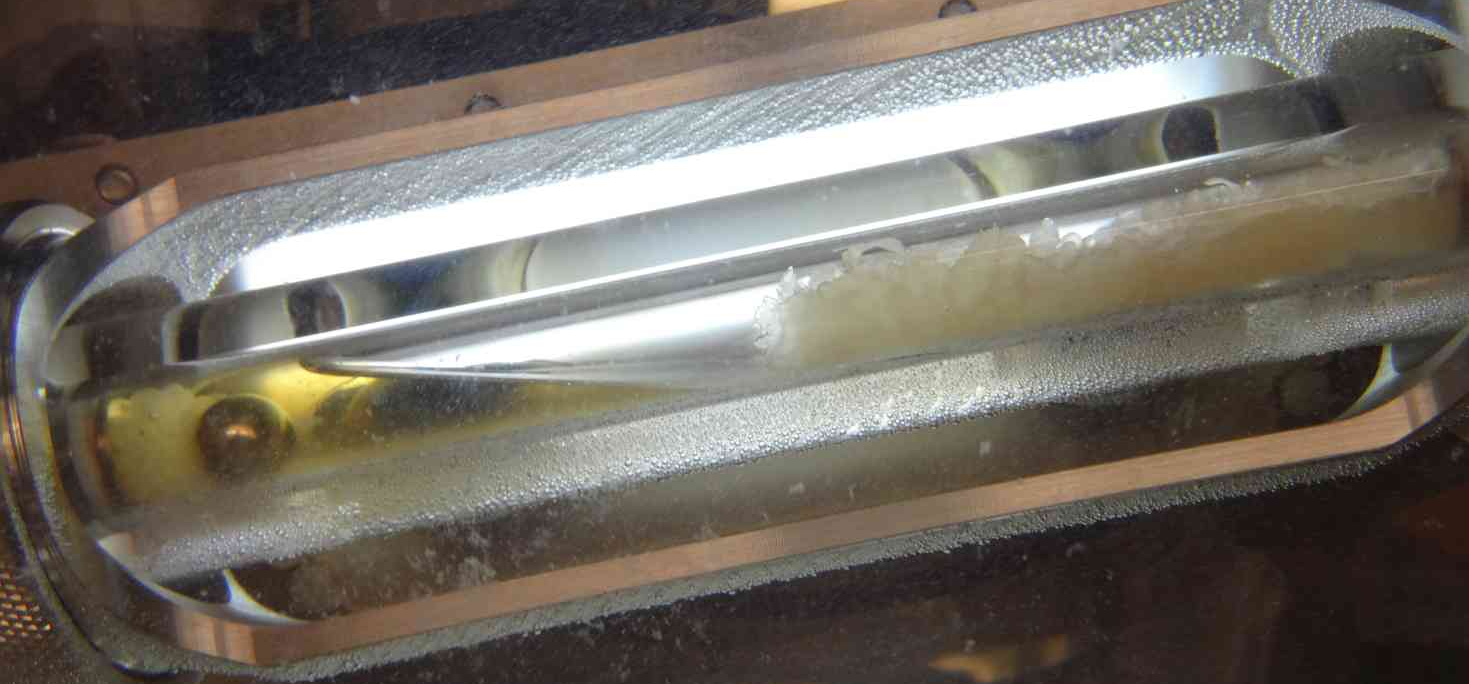}
  \caption{Clear hydrate plugs.}
  \label{fig:picture_cell_dose05}
 \end{subfigure}
 \caption{Examples of progressive hydrate formation as the AA dose rate is reduced from high (\ref{fig:picture_cell_dose20}) to low (\ref{fig:picture_cell_dose05}).}
 \label{fig:picture_cell}
\end{figure}

\begin{figure}
 \includegraphics[width=1.00\textwidth]{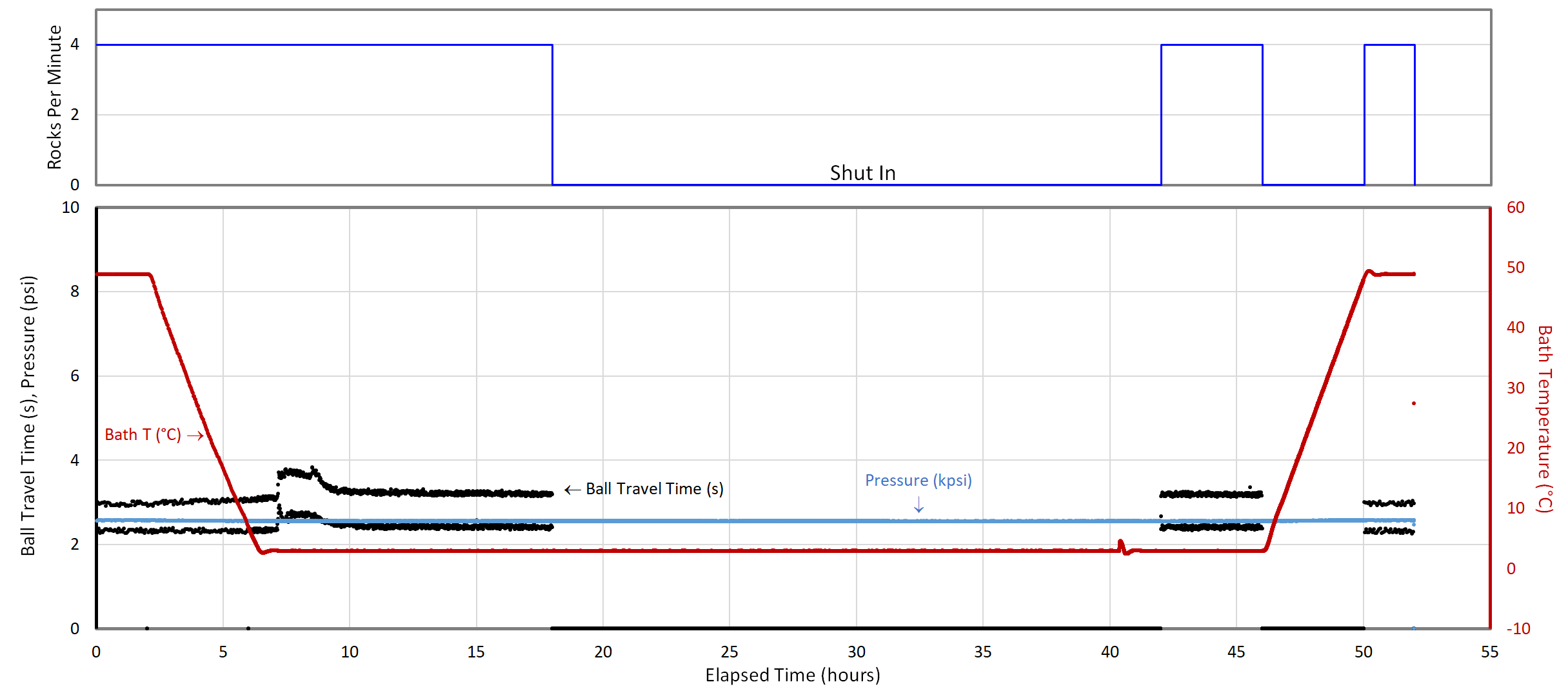}
 
 \vspace{30pt}
 
 \includegraphics[width=1.00\textwidth]{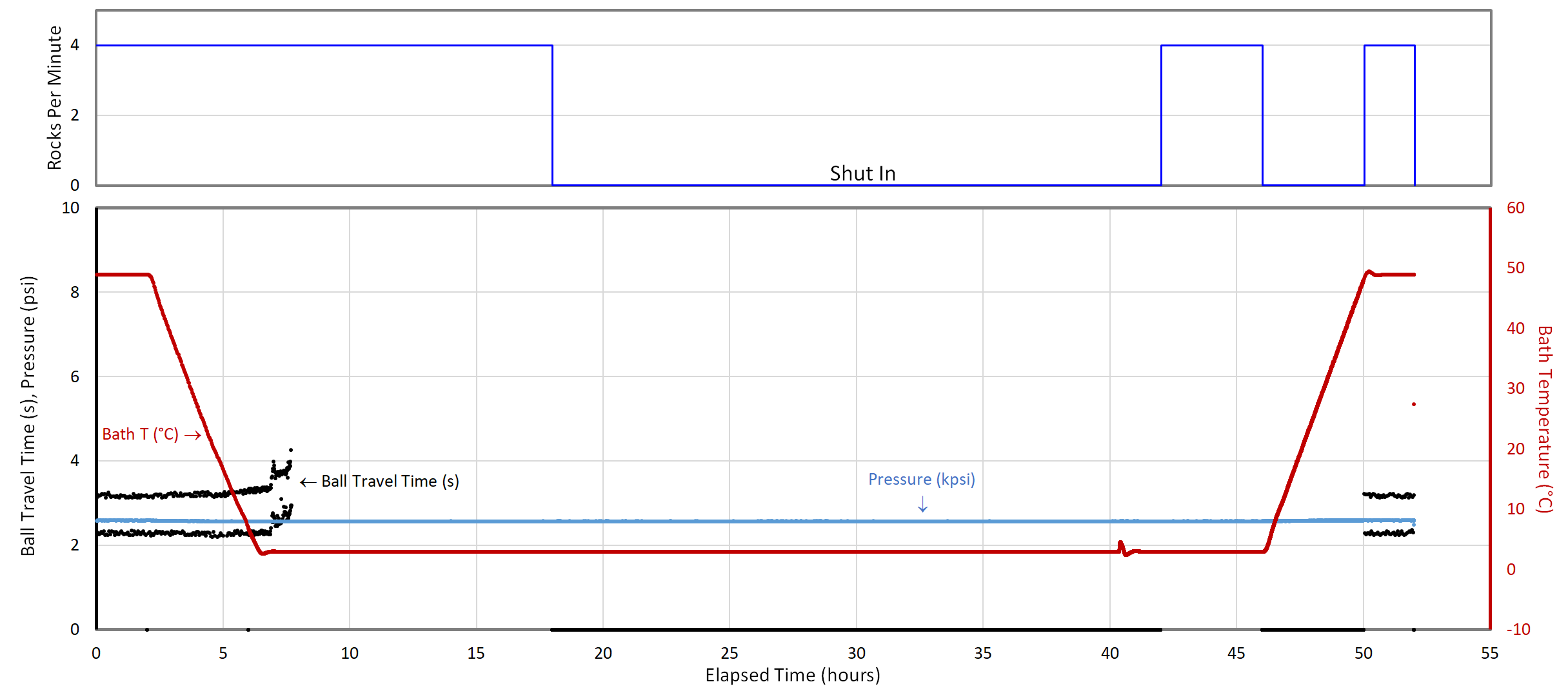}
 \caption{Two illustrative examples of the ball travel times from the rocking cell experiments, showing a clear pass (top) and fail (bottom). In the case of the pass, the ball keeps moving until the shut-in step even though the temperature is reduced, and recovers the movements once the rocking is switched on again. In the case of the fail, on the other hand, the ball gets blocked and stops moving (after roughly 8 hours) due to the formation of hydrate plugs, and only recovers the movements once the temperature is increased again. The apparently two parallel lines are merely an artifact, since the measurement of the travel time into one direction is slightly delayed with respect to the measurement into the other direction.}
 \label{fig:ball_times}
\end{figure}

To determine the relative performance of each inhibitor or dose rate of inhibitor, visual observations were made during the shut in period and correlated with an interpretation of the time required for the ball within the cell to travel between two magnetic sensors.
In Figure~\ref{fig:picture_cell} we show examples of the progressive hydrate formation as the dose rate is reduced, and in Figure~\ref{fig:ball_times} two illustrative examples of the ball travel time indicating a clear pass and a clear fail.
Each experiment was conducted in duplicate to confirm reproducibility.

%
%

\section{Results}

\subsection{Non-steered simulations}

We initially carried out 24 non-steered simulations up to \SI{600}{ns} for the 4 different AA systems at 3 different surface concentrations with the water droplet either covered or not covered with the AA molecules at the same concentration as that of the hydrate surface. We will begin the presentation of the results with a qualitative description of the simulations based on visual inspection of the trajectories that was carried out on all the runs.

\paragraph{Coalescence process}

In a typical trajectory, the water droplet, covered or not covered with AA molecules, diffuses inside the liquid hydrocarbon phase. During its passage, the droplet can move in and out of proximity to the AA-covered hydrate surface multiple times. At a certain moment, the droplet can irreversibly coalesce with the hydrate surface. The average time that the droplet can freely diffuse inside the hydrocarbon phase depends on the type of the AA and the surface concentration although it should be kept in mind that this is an inherently stochastic process, and thus the time required for coalescence (henceforth called coalescence time) can significantly vary even for identical systems. It should be noted that the presence of AAs on the water droplet only slightly reduces the diffusion coefficient of the center of mass of the droplet as a function of the concentration, as shown below. During coalescence, a capillary water bridge can be observed, especially at low surface concentrations. The coalescence proceeds with the wetting of the hydrate surface, and given enough time the water molecules of the droplet spread evenly on the hydrate surface. This wetting time also depends on the AA concentration as the droplet water molecules are required to pass through the AA film on the hydrate surface.

\paragraph{Coalescence time}

The first striking difference is observed between the systems where the water droplet is not covered with AAs versus the systems where the droplet is covered. In almost all the cases for the not covered droplet systems, coalescence does take place even at the high surface concentration examined here. Additionally, the coalescence time for the not covered systems is significantly lower than in the case where the droplet is covered. This shows that preventing agglomeration between a hydrate particle and a water droplet requires the water droplet to be covered with AAs. Finally, the coalescence time for the not covered systems increases as the surface concentration increases.

In the systems where the water droplet is covered, differences are observed between the low, medium and high concentrations. In medium and high concentrations scenarios, no coalescence was observed for all the 4 different AA systems examined. We note again that these simulations extended up to \SI{600}{ns}, indicating that all surfactants examined can act as an AA at high surface concentrations. In the case of low concentrations, there were differences between the different AA systems; for some AAs coalescence did happen, while for others it did not. Evidently, this is the most interesting concentration, and thus we carried out multiple independent runs for each system at low surface concentration so our results obtain a greater statistical significance. The results of these first 24 runs are presented in Table~\ref{tab:table_non_steered_runs}.

\begin{table}[H]
\centering
\bgroup
\def\arraystretch{1.5}
\resizebox{\textwidth}{!}{%
\begin{tabular}{ccccccccccccc}
\hline
System                                         & \multicolumn{3}{c}{\textbf{AA1}}                                                                                                                        & \multicolumn{3}{c}{\textbf{AA2}}                                                                                                                        & \multicolumn{3}{c}{\textbf{AA3}}                                                                                                                        & \multicolumn{3}{c}{\textbf{AA4}}                                                                                                   \\
\multicolumn{1}{c|}{\textbf{}}                 & \begin{tabular}[c]{@{}c@{}}Time \\ {[}ns{]}\end{tabular} & Result & \multicolumn{1}{c|}{\begin{tabular}[c]{@{}c@{}}Coal. time \\ {[}ns{]}\end{tabular}} & \begin{tabular}[c]{@{}c@{}}Time \\ {[}ns{]}\end{tabular} & Result & \multicolumn{1}{c|}{\begin{tabular}[c]{@{}c@{}}Coal. time \\ {[}ns{]}\end{tabular}} & \begin{tabular}[c]{@{}c@{}}Time \\ {[}ns{]}\end{tabular} & Result & \multicolumn{1}{c|}{\begin{tabular}[c]{@{}c@{}}Coal. time \\ {[}ns{]}\end{tabular}} & \begin{tabular}[c]{@{}c@{}}Time \\ {[}ns{]}\end{tabular} & Result & \begin{tabular}[c]{@{}c@{}}Coal. time \\ {[}ns{]}\end{tabular} \\
\multicolumn{1}{c|}{\textbf{uncovered low}}    & 200                                                      & coal   & \multicolumn{1}{c|}{22}                                                             & 200                                                      & coal   & \multicolumn{1}{c|}{178}                                                            & 200                                                      & coal   & \multicolumn{1}{c|}{7}                                                              & 200                                                      & coal   & 36                                                             \\
\multicolumn{1}{c|}{\textbf{uncovered medium}} & 200                                                      & coal   & \multicolumn{1}{c|}{40}                                                             & 200                                                      & coal   & \multicolumn{1}{c|}{92}                                                             & 200                                                      & coal   & \multicolumn{1}{c|}{60}                                                             & 200                                                      & coal   & 44                                                             \\
\multicolumn{1}{c|}{\textbf{uncovered high}}   & 300                                                      & coal   & \multicolumn{1}{c|}{135}                                                            & 600                                                      & inhib  & \multicolumn{1}{c|}{n/a}                                                            & 200                                                      & coal   & \multicolumn{1}{c|}{54}                                                             & 600                                                      & inhib  & n/a                                                            \\
\multicolumn{1}{c|}{\textbf{covered low}}      & 600                                                      & inhib  & \multicolumn{1}{c|}{n/a}                                                            & 600                                                      & inhib  & \multicolumn{1}{c|}{n/a}                                                            & 200                                                      & coal   & \multicolumn{1}{c|}{90}                                                             & 200                                                      & coal   & 138                                                            \\
\multicolumn{1}{c|}{\textbf{covered medium}}   & 600                                                      & inhib  & \multicolumn{1}{c|}{n/a}                                                            & 600                                                      & inhib  & \multicolumn{1}{c|}{n/a}                                                            & 600                                                      & inhib  & \multicolumn{1}{c|}{n/a}                                                            & 600                                                      & inhib  & n/a                                                            \\
\multicolumn{1}{c|}{\textbf{covered high}}     & 600                                                      & inhib  & \multicolumn{1}{c|}{n/a}                                                            & 600                                                      & inhib  & \multicolumn{1}{c|}{n/a}                                                            & 600                                                      & inhib  & \multicolumn{1}{c|}{n/a}                                                            & 600                                                      & inhib  & n/a                                                            \\ \hline
\end{tabular}%
}
\egroup
\caption{Results of the non-steered MD runs to study the coalesence between the hydrate slab and the water droplet, for all four AAs and the six different setups. The simulations were stopped at 600 ns, or earlier if coalescence has occurred.}
\label{tab:table_non_steered_runs}
\end{table}

\begin{figure}
 \includegraphics[width=1.0\textwidth]{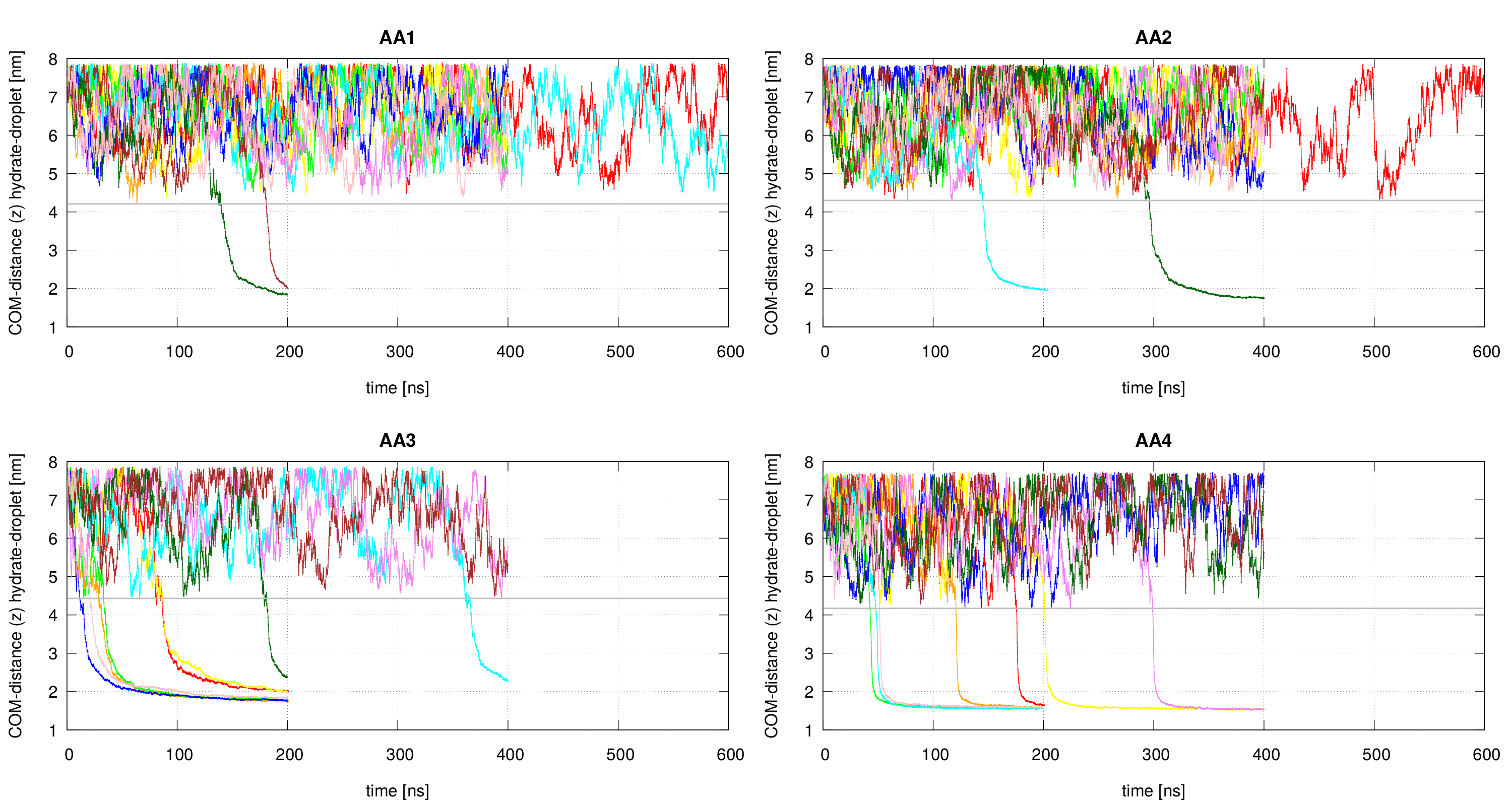}
 \caption{Distance between the centers of masses of the hydrate slab and the water droplet. The sharp decreases indicate coalescence. For each AA, 10 independent runs were performed. The gray line indicates the boundary beyond which the irreversible coalescence starts.}
 \label{fig:figure_non_steered_runs}
\end{figure}

\begin{figure}
 \includegraphics[width=0.5\textwidth]{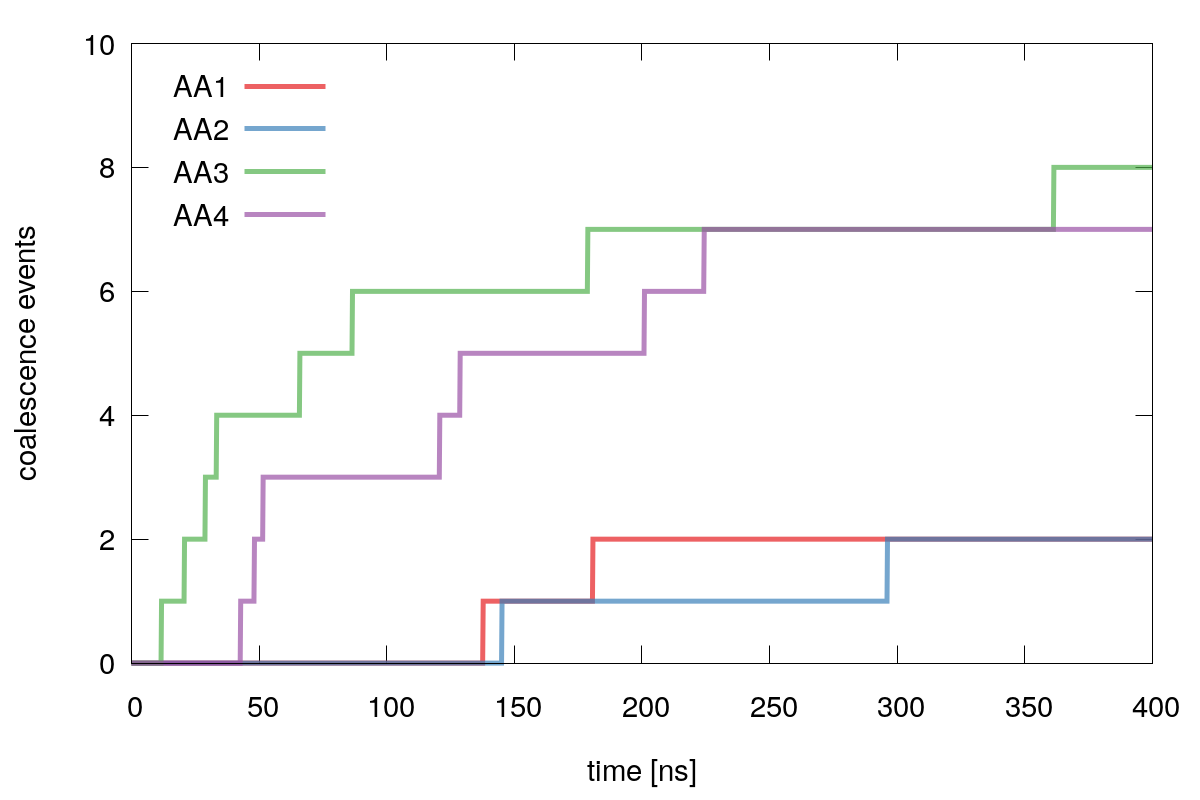}
 \caption{Number of coalescence events as a function of the simulation time, for the data of Figure~\ref{fig:figure_non_steered_runs}.}
 \label{fig:agglomeration_events}
\end{figure}

For each of the 4 AA systems at low AA surface concentration and with the water droplet being covered, we carried out 10 independent runs up to 400 ns. By counting how many times coalescence was observed for each AA molecule a ranking can be obtained. The results indicate that AA1 and AA2 are the best performing molecules, as for both AAs coalescence only took place 2 out of 10 times in the independent runs (2/10). Worse performance was obtained for AA4 (7/10), followed closely by AA3 (8/10). The results are presented in Figure~\ref{fig:figure_non_steered_runs}. It can also be observed that the coalescence time varies significantly for each system. A greater sample, both in number of runs and time, would in principle provide a clearer distinction between the performance of the AAs, but the purpose here is to demonstrate, within reasonable use of computational resources, the capability of brute force simulations to provide a qualitative description of the AA performance of the different surfactants.
Indeed, we see from Figure~\ref{fig:agglomeration_events}, showing the number of coalescence events as a function of the simulation time, that the chosen simulation time of \SI{400}{ns} is clearly enough to distinguish between AA1 and AA2 on the one hand and AA3 and AA4 on the other hand.

\paragraph{Density profiles}

\begin{figure}
 \begin{subfigure}{0.40\textwidth}
  \includegraphics[width=1.00\textwidth]{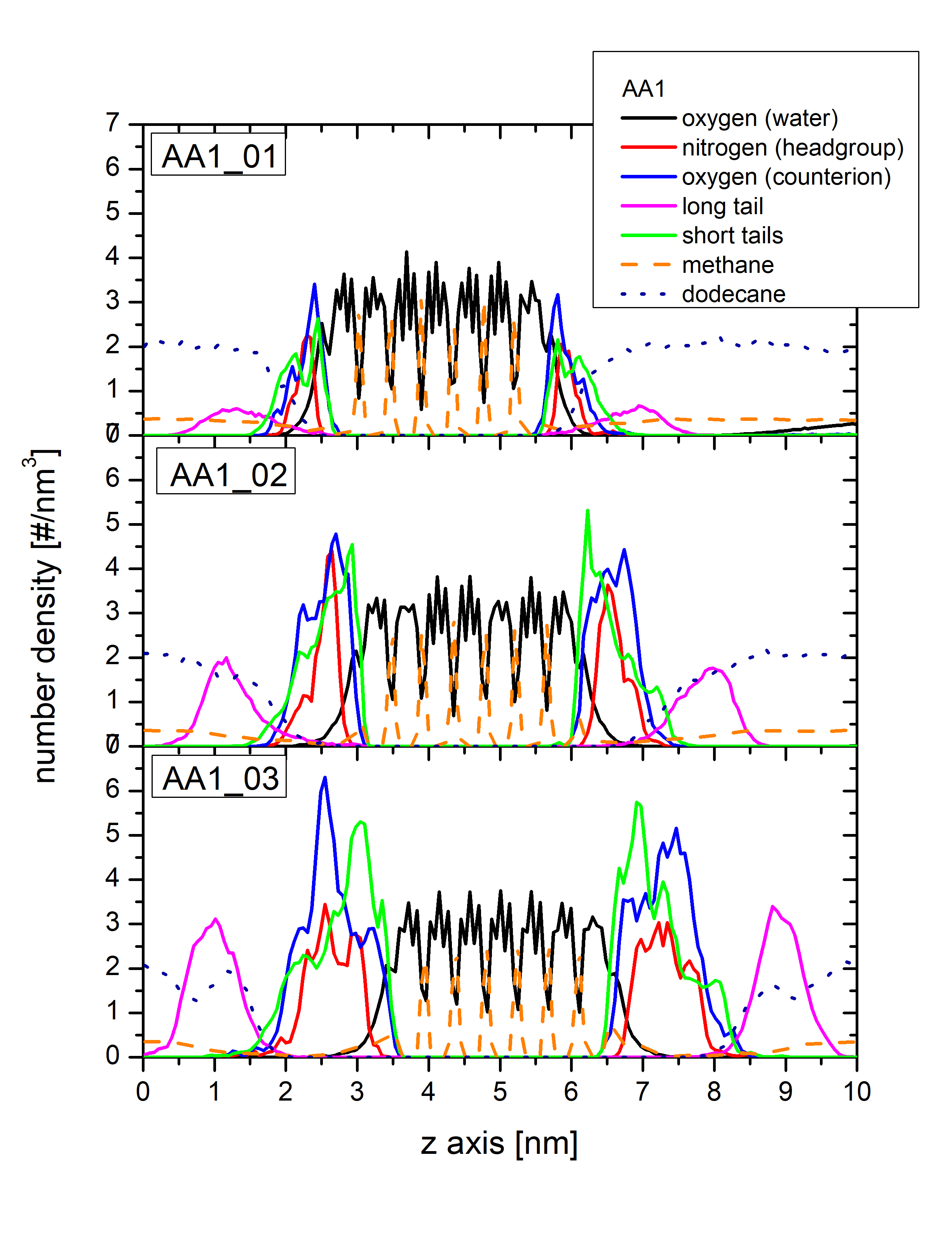}
  \caption{AA1}
  \label{fig:density_profiles_AA1}
 \end{subfigure}
 \begin{subfigure}{0.40\textwidth}
  \includegraphics[width=1.00\textwidth]{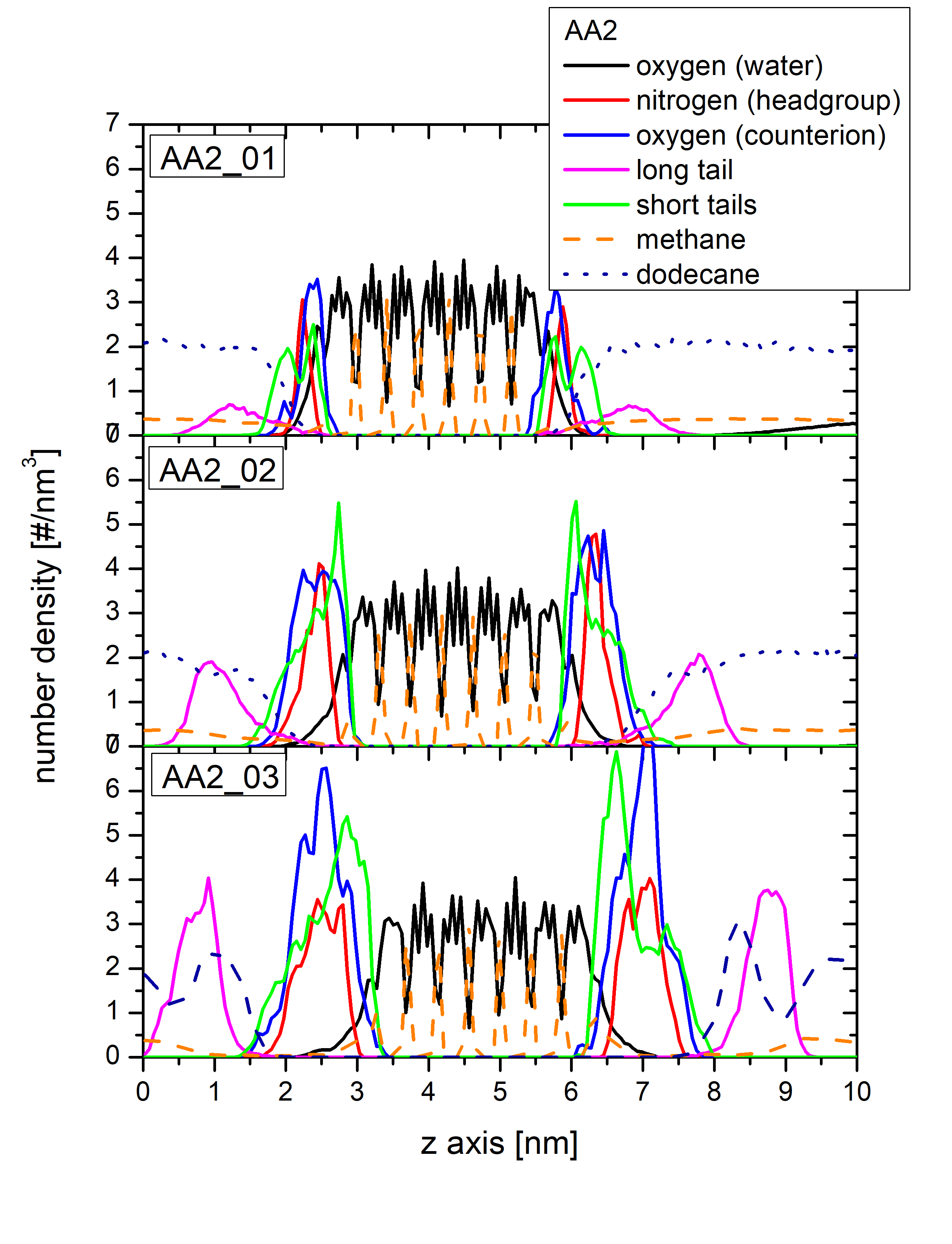}
  \caption{AA2}
  \label{fig:density_profiles_AA2}
 \end{subfigure}
 \begin{subfigure}{0.40\textwidth}
  \includegraphics[width=1.00\textwidth]{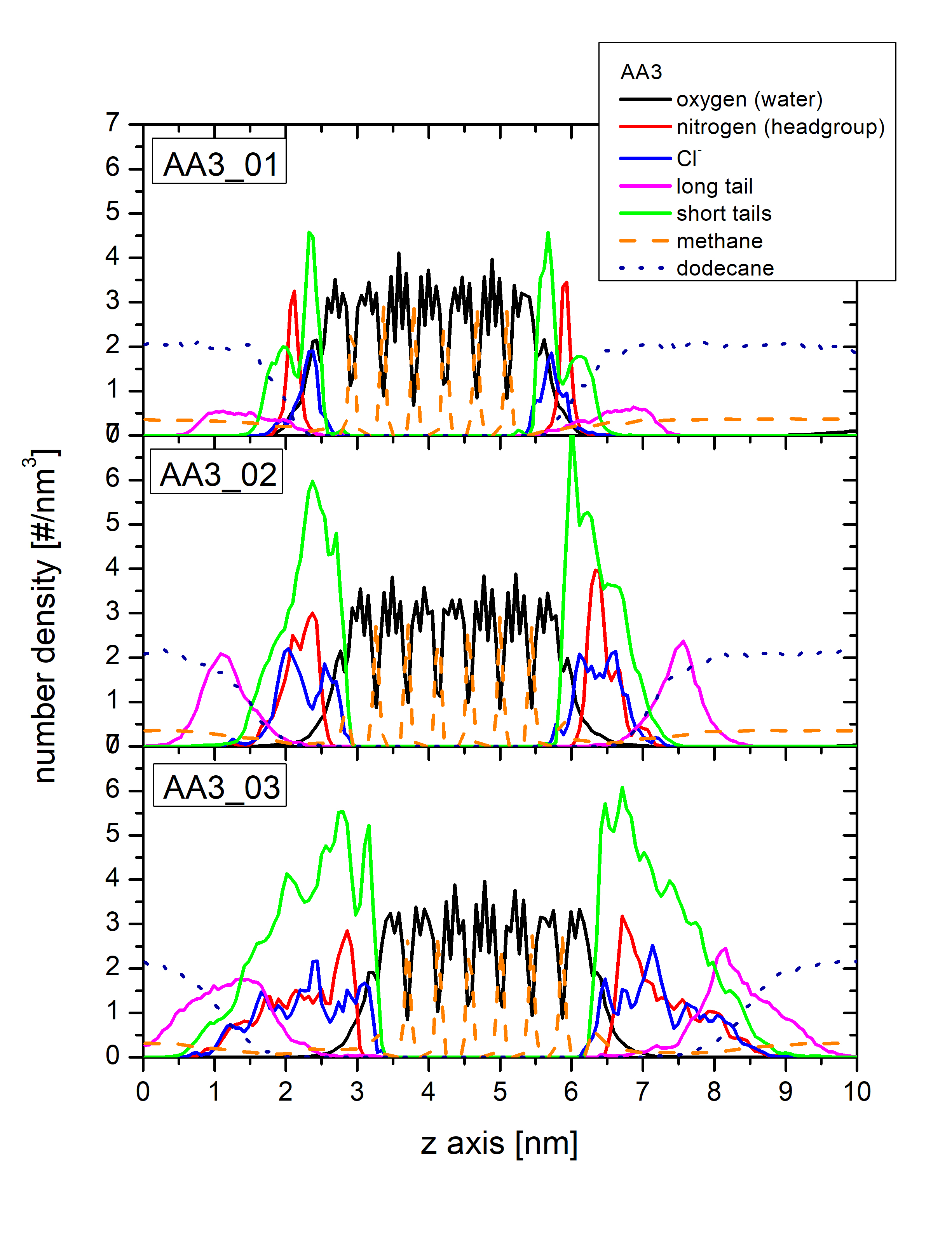}
  \caption{AA3}
  \label{fig:density_profiles_AA3}
 \end{subfigure}
 \begin{subfigure}{0.40\textwidth}
  \includegraphics[width=1.00\textwidth]{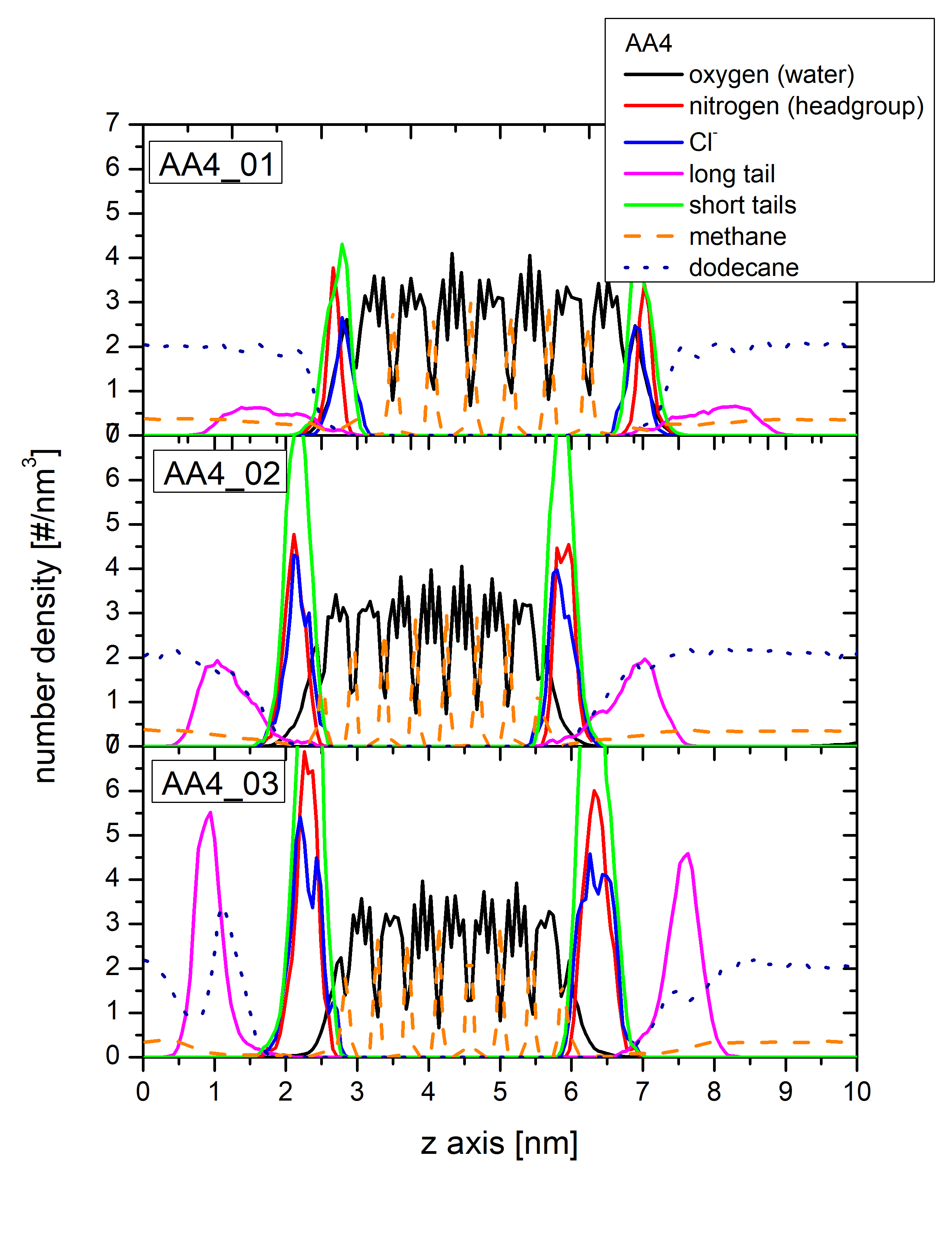}
  \caption{AA4}
  \label{fig:density_profiles_AA4}
 \end{subfigure}
 \caption{Number density profiles along the normal to the hydrate surface axis, for (a) AA1, (b) AA2, (c) AA3, and (d) AA4. We show the results for all three AA concentrations, namely low (top), medium (middle) and high (bottom). The species presented are: oxygen atoms of the waters (black line), nitrogen atoms of the AA headgroup (red line), oxygens/chlorides of the counterion (blue line), terminal carbon atom of the long tail (purple line), terminal carbon atom of the short tails (green line), methane (dashed orange line), and n-dodecane (dotted royal blue line).}
 \label{fig:density_profiles}
\end{figure}

Next, a series of density profiles of the AA films on the hydrate surface is presented for each of the 4 different AAs at the low, medium and high surface concentration. The results are shown in Figure~\ref{fig:density_profiles}. The number density as a function of their position along the z-axis, which is normal to the hydrate surface, is given for a number of species. For reasons of clarity, the number density of the water oxygens and methane molecules is scaled down by a factor of 10 and 5, respectively.

In Figure~\ref{fig:density_profiles_AA1} we show the results for AA1. The number density profile of the oxygen atoms of the water reveals the position of the hydrate crystal, with a film of AA on each side. It also shows the existence of a liquid water layer on the surface of the hydrate, which can also be observed visually in the simulation trajectories. Thus, the pre-melting phenomenon is successfully captured. Both the long and short tails of the AA molecules are represented by the density profiles of the terminal carbon atoms of these tails. The long tails are found, as expected, farther away from the hydrate surface, but for the low and medium concentrations a few long tails can be found close to the surface, revealing long tail orientations not normal to the surface. In the high concentration case this is not happening, and additionally the distance of the mean of the distribution from the hydrate surface is about 0.5 nm larger than in the other two cases, showing that this concentration is quite high and there is not enough room for all the AA molecules to be adsorbed on the surface. This observation is further supported by the distribution of the short tails, which is found very close to the surface and is skewed towards it. This indicates that most of the short tails are adsorbed on the surface, but not all of them. Equivalently close to the surface are the oxygens of the acrylate counterion, while only slightly further away is the position of the nitrogen atom, which indicates the position of the headgroup. It is evident that the cationic headgroup interacts strongly with the negatively charged counterion, both closely positioned on the hydrate surface. The profile of the methane molecules reveals that methane is generally excluded from the AA film, primarily close to the headgroup, with this effect being more pronounced as the surface concentration increases. Finally, the position of the n-dodecane molecules of the hydrocarbon phase is shown through the number density of the 6th atom of the hydrocarbon chain. It can be observed that the n-dodecane molecules participate in the AA film, and are progressively excluded from being in the proximity to the hydrate surface as the AA concentration increases.

The number density profiles for AA2 are shown in Figure~\ref{fig:density_profiles_AA2} and are in almost every aspect similar to the AA1 profiles. These two surfactants differ only in their counterion, which does not seem to affect the AA film on the hydrate surface. 

In Figure~\ref{fig:density_profiles_AA3} similar number density profiles are presented for AA3. In comparison with the AA1 case at low concentrations, the distribution of the nitrogen of the headgroups is narrower and slightly farther away from the surface than the chlorine counterion. The long tail profiles are very similar. At higher concentrations, both the long and short tails distributions are much broader, indicating that the AA3 films are less ordered at these concentrations, most probably due to steric hindrances as the AA3 molecule has three short tails while the AA1 molecule has two.

On the other hand, the density profiles of AA4, which are presented in Figure~\ref{fig:density_profiles_AA4}, show a much more ordered AA film especially for the higher concentration. Again, both the positively charged headgroup, as shown by the nitrogen profile, and the chlorine anion are very close to the surface with their average distance from it being almost the same. The distribution of the long tails also reveals the existence of a well-ordered AA film.

\paragraph{Droplet Diffusion}

\begin{figure}
 \includegraphics[width=0.9\textwidth]{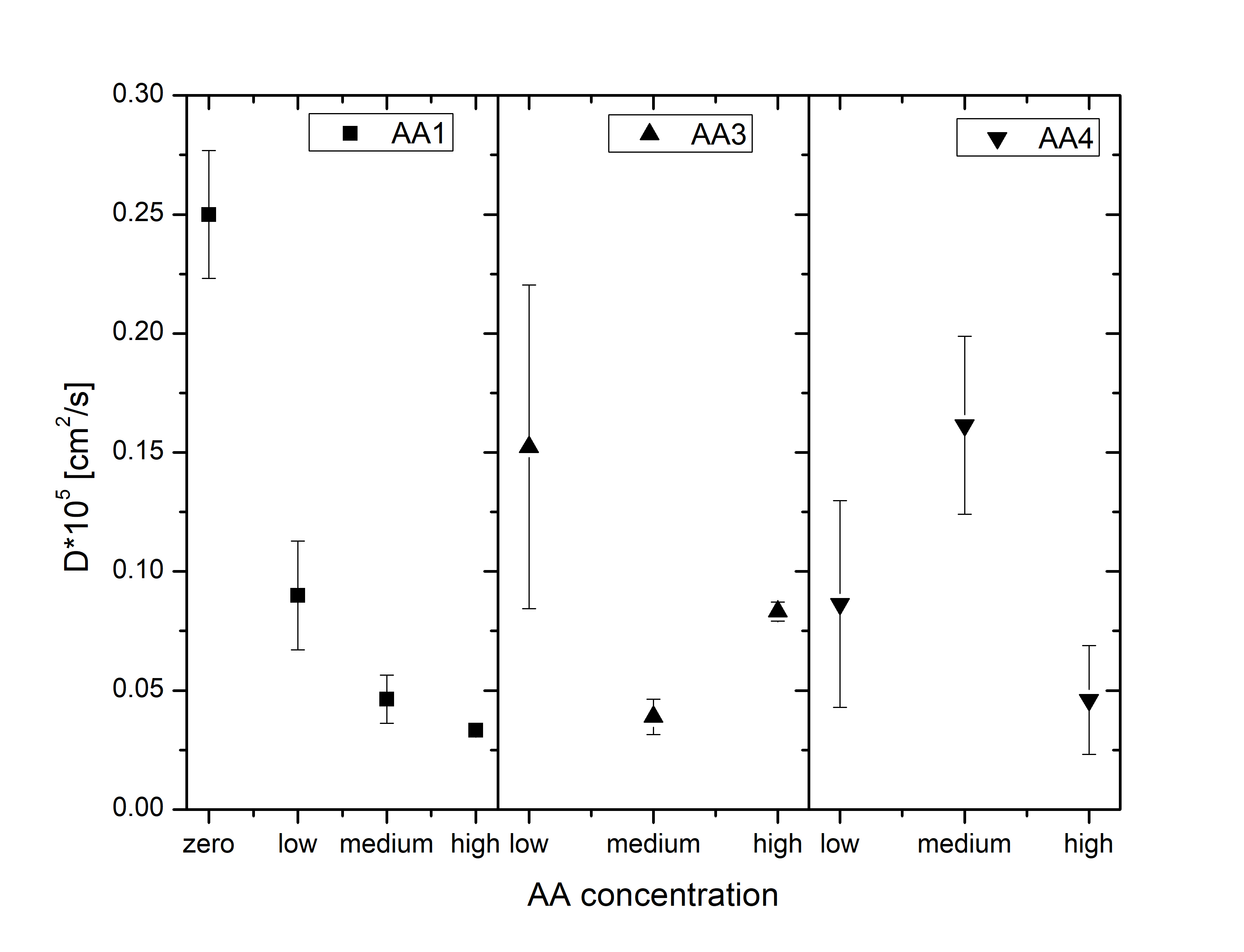}
 \caption{Water droplet diffusion coefficient inside the liquid hydrocarbon phase, covered with the AA1 (and by inference AA2), AA3 and AA4 molecules. The uncovered case is also presented.}
 \label{fig:diffusion-water-droplet}
\end{figure}

The diffusion coefficient of the water droplet inside the hydrocarbon phase as a function of the droplet AA surface concentration is presented in Figure~\ref{fig:diffusion-water-droplet}. The uncovered case is also included. As can be seen, covering the droplet reduces diffusion, but there is no significant trend correlating the AA concentration with the diffusion coefficient. More interestingly, the diffusion coefficients for each surface concentration and AA type are very similar, providing a basis for the comparison of the behavior of the different AA systems. The calculation of the diffusion coefficient of AA2 was omitted as it is expected to be the same as that of AA1. 

\paragraph{Lateral diffusion on the surface}

\begin{figure}
 \includegraphics[width=0.5\textwidth]{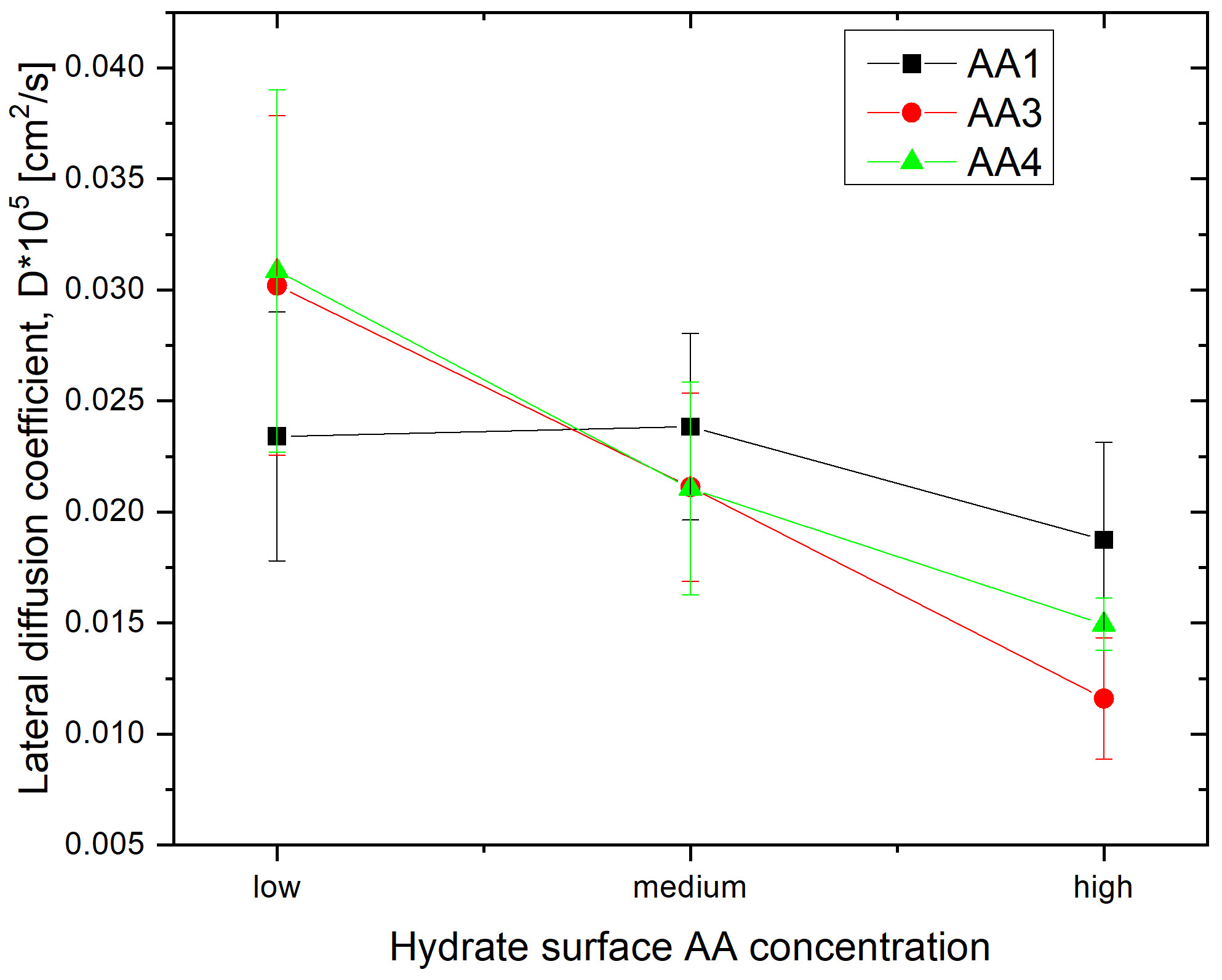}
 \caption{Lateral diffusion coefficient of the nitrogen atom of the headgroup of the various AAs on the hydrate surface.}
 \label{fig:diffusion-lateral}
\end{figure}

In Figure~\ref{fig:diffusion-lateral} the lateral diffusion coefficients of the nitrogen atom of the headgroup of the AA molecules on the hydrate surface as a function of the surface concentration are presented. It can be inferred that there are no significant differences in the mobility of the headgroups of the examined AAs composing the film on the hydrate surface. Nevertheless, these results do not reflect the behavior of the AA upon the coalescence procedure, but only provide insight into the ``static'' behavior of these AA films.

\paragraph{Long tail orientation}

\begin{figure}
 \includegraphics[width=0.7\textwidth]{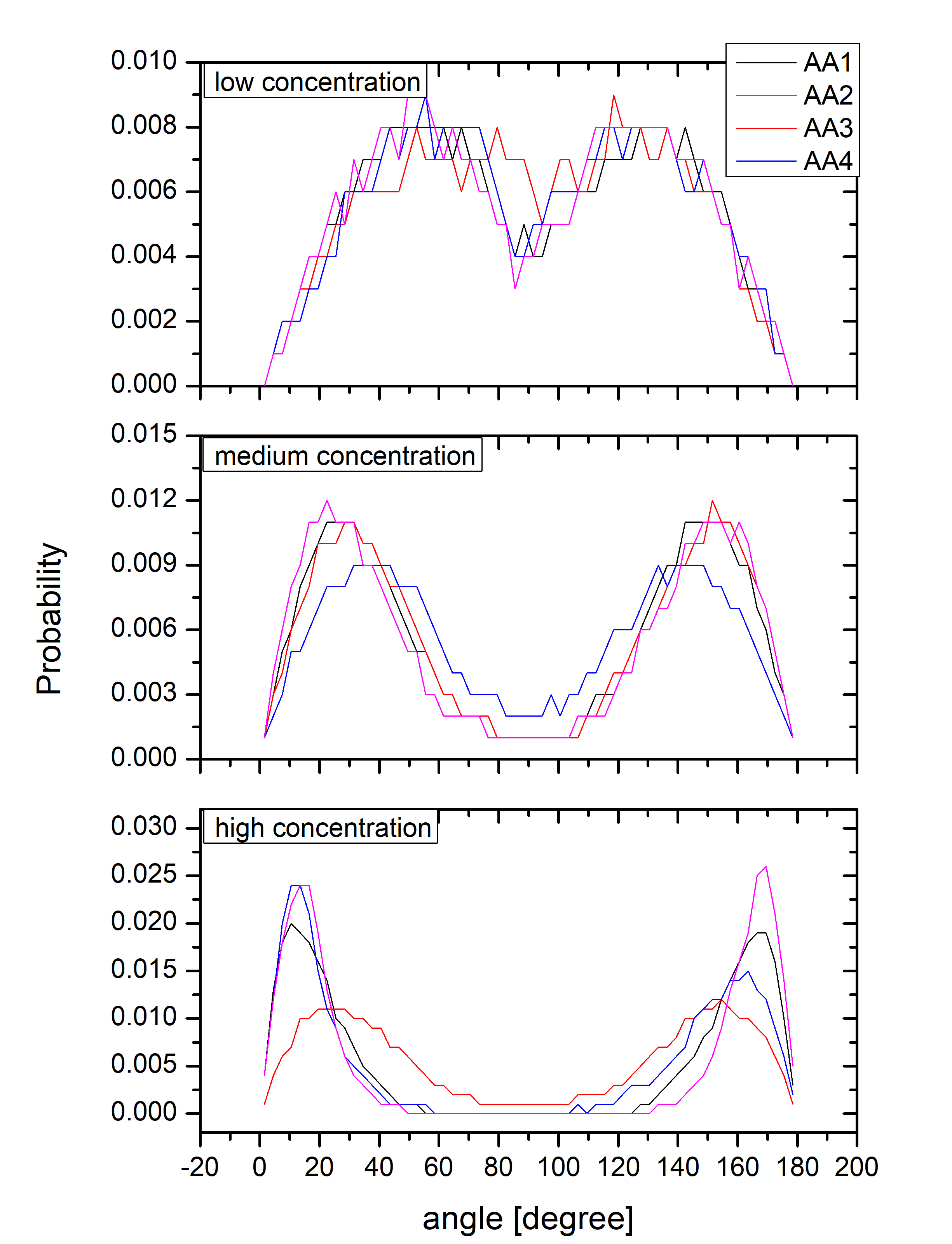}
 \caption{Probability distributions of the angle formed between the normal to the hydrate surface and the vector defined by the nitrogen atom of the headgroup and the terminal atom of each long tail.}
 \label{fig:tail_orientations}
\end{figure}

The orientation of the long tails for each AA molecule at all concentrations was further examined through the calculation of the distribution of the angle between the normal to the hydrate surface and the vector defined by the nitrogen atom of the headgroup and the terminal atom of each long tail. The results are presented in Figure~\ref{fig:tail_orientations}. The closer the angle is to the values 0 and 180 degrees the more perpendicular to the hydrate surface is the orientation of the long tails, corresponding to the right and left side of the hydrate slab, respectively. It can be observed that at low concentration the average orientation is 45 degrees while some long tails are parallel to the surface. As the concentration increases the orientation becomes more perpendicular to the surface and at the highest concentration the film order is the highest. Exception to that is the case for AA3, whose film is less ordered at the highest surface concentration, in accordance with the results of the density profiles.

\subsection{Steered simulations}

\begin{figure}
 \includegraphics[width=1.0\textwidth]{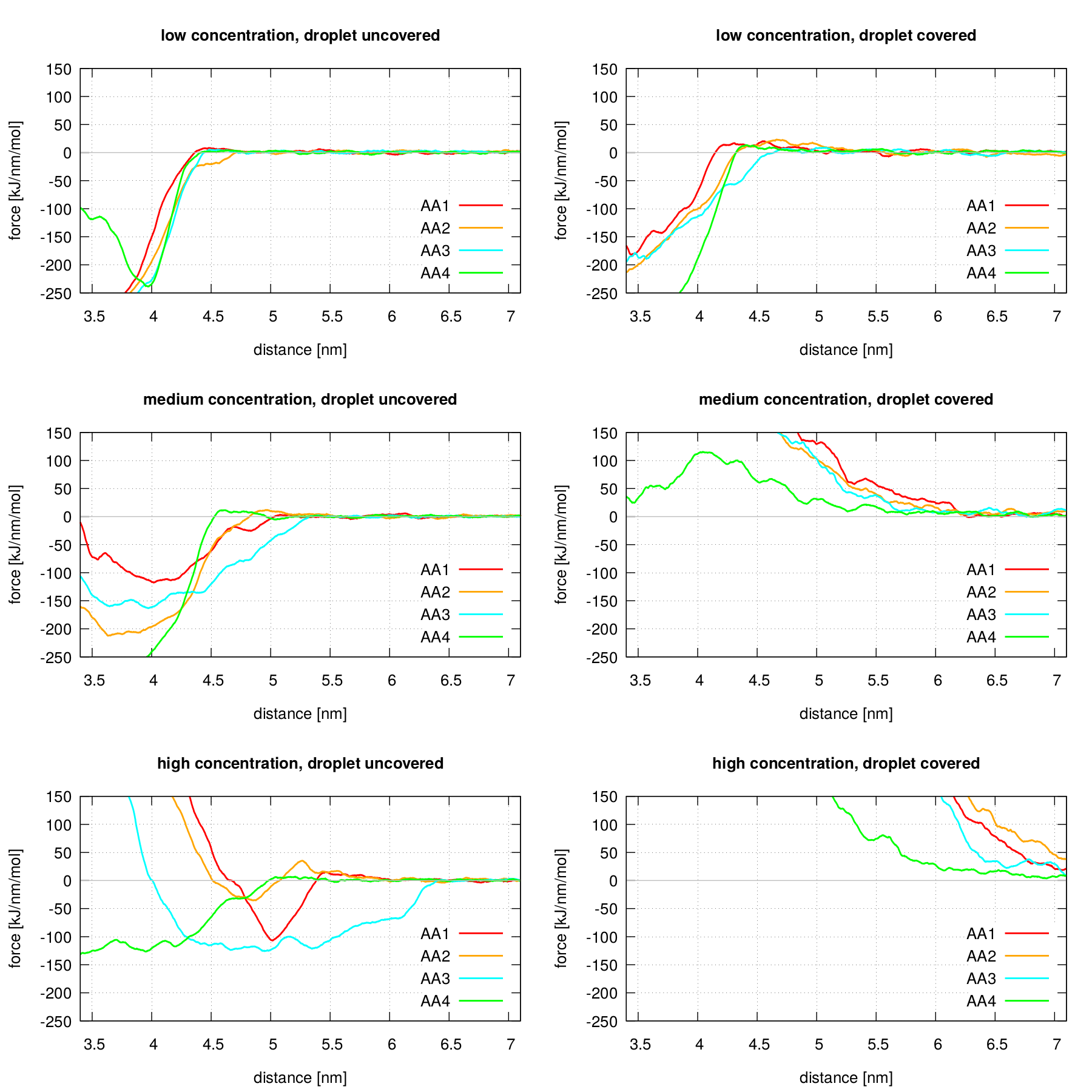}
 \caption{Force-distance profiles for each of the fours AAs and six setups. For each of the 24 systems, 5 independent runs were preformed; the plot shows the resulting average curves. The indicated distance is the z component of the distance between the center of mass of the droplet and a reference position within the hydrate slab.}
 \label{fig:distance-force_all}
\end{figure}

We performed steered pulling simulations for all the 24 mentioned setups, pulling the water droplet towards the hydrate surface. Despite the steered movement of the water molecules of the droplet, the simulations still contain a considerable amount of stochasticity since the AA molecules can move freely. This is important, since the interactions between the AAs on the hydrate surface and the (covered) droplet are fundamental for coalescence inhibition. Therefore, we performed 5 independent runs for each of the 24 simulation setups, and calculated the average force-distance profiles, which are presented in Figure~\ref{fig:distance-force_all}.

As can be seen from these plots, the force is zero at large separations between the hydrate surface and the water droplet, indicating that there is no significant interaction between surface and droplet. As the droplet gets closer, it starts to interact with the AA molecules on the hydrate surface. In the case where the droplet is as well covered with AAs, the interaction takes place firstly among the AAs of the surface and the AAs of the droplet. This first interaction is generally repulsive, as shown by the increase of the force which is required to keep the water droplet moving with a constant velocity towards the hydrate surface, and it depends on the nature of the AA molecules and their concentration. Once this barrier has been crossed, the water molecules come close enough to the hydrate surface and the coalescence process proceeds via the formation of a liquid water bridge. This stage is strongly favored and therefore leads to a pronounced region of negative force. When the droplet is pulled even further, it starts to collide with the hydrate surface, and consequently the force increases strongly. However, this part of the force-distance profile does not represent any more a meaningful process and should be ignored. The part of the force-distance profile that better reflects the inhibition potency of an AA molecule and therefore represents the most interesting region is the one where the interaction of the (coated) droplet with the surface AAs begins.

Independently of the specific AA molecule, two more observations can be made. Firstly, the interactions become more repulsive the higher the AA concentration is. Secondly, the interactions are more repulsive for the covered droplet compared to the uncovered droplet. These observations are in line with the behavior observed in the non-steered simulations. Regarding the differences between the four AAs, a clear ordering of the molecules across all setups is difficult, but overall AA1 and AA2 seem to be more repulsive compared to AA3 and AA4.

Although the force-distance profiles can offer a qualitative description of the interactions between the hydrate surface and the droplet for each of the various AA setups, it would be interesting to describe these interactions, especially the initial repulsive ones, more quantitatively. The external work exerted onto a system to bring it from one equilibrium state to another through a non-equilibrium process, $\Delta W$, can be related to the free energy difference between the two equilibrium states, $\Delta E$, using the Jarzynski equality~\cite{Jarzynski1997-Nonequilibrium}:

\begin{equation}
 e^{-\beta\Delta E} = \langle e^{-\beta\Delta W} \rangle \,.
\end{equation}

In the case of steered molecular dynamics, a harmonic bias potential is added to the original Hamiltonian, and the Jarzynski equality relates the external work required to move the center of the harmonic bias at a constant rate, $\Delta W_B$, with the free energy change of this biased system, $\Delta E_B$, i.e.

\begin{equation}
  e^{-\beta\Delta E_B} = \langle e^{-\beta\Delta W_B} \rangle \,.
\end{equation}

The external work can be calculated by integrating the force exerted by the bias potential, $F_B$, onto the collective variable describing the transition between the two states. In our case, the initial state $a$ corresponds to the solvated droplet far from the surface, whereas the final state $b$ corresponds to the droplet ``close'' to the surface (exact definitions below). Considering as collective variable the distance $z$ to the surface we obtain

\begin{equation}
 \Delta W_B = W_B(b) - W_B(a) = -\int_a^b F_B(z)\,\mathrm{d}z \,.
\end{equation}

Finally, using a first approximation to the Jarzynski equality, we obtain the free energy difference between the two states by averaging over all the trajectories:

\begin{equation}
 \Delta E_B \approx \langle \Delta W_B \rangle = -\int_{a}^{b} \langle F_B(z) \rangle \,\mathrm{d}z \,.
 \label{eq:free_energy_difference}
\end{equation}

As was shown by Park and Schulten~\cite{Park2004-Calculating}, the free energy change of the biased system is the same as that of the unbiased one, i.e. $\Delta E_B \approx \Delta E$, if the harmonic bias is strong enough, being called the "stiff-spring approximation".

As in the case of the non-steered runs, we focused on the case where both surface and droplet are covered at low concentration. To determine good integration bounds $a$ and $b$, we again analyzed the center of mass distance between the hydrate surface and the droplet for the non-steered simulations. These distances, plotted as a function of the simulation time, are shown in Figure~\ref{fig:figure_non_steered_runs} for the low surface concentration case. Based on the minimum distance where no coalescence took place throughout those trajectories, we can define a critical value of this distance that seems to be a ``decision boundary''. If the distance becomes smaller than this critical value, coalescence is unavoidable. We therefore chose this critical value to be the upper integration bound (i.e.\ state $b$), yielding a value of \SI{4.21}{nm} for AA1, \SI{4.30}{nm} for AA2, \SI{4.43}{nm} for AA3 and \SI{4.17}{nm} for AA4. The lower integration bound (i.e.\ state $a$) was set to \SI{6.5}{nm}. 

\begin{figure}
 \includegraphics[width=0.5\textwidth]{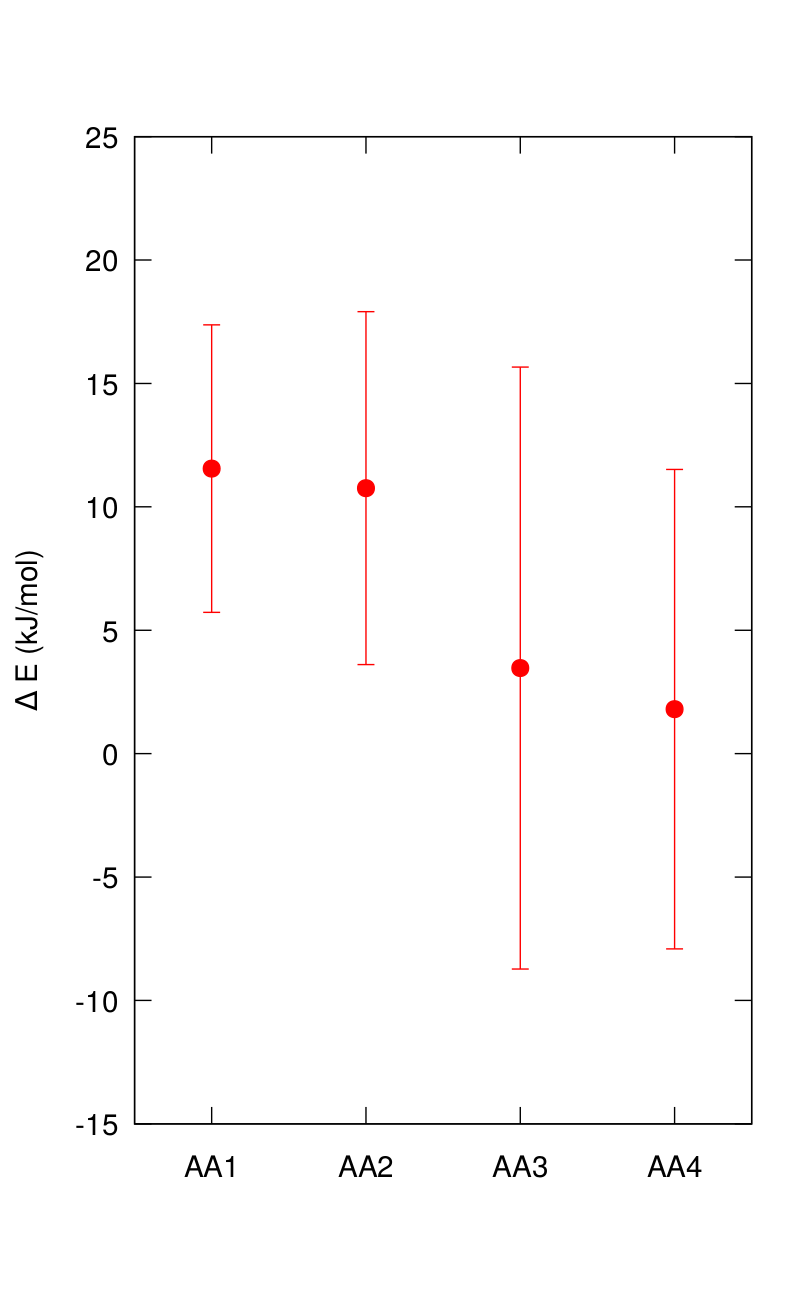}
 \caption{Approximate free energy barriers for the coalescence between the water droplet and the hydrate slab, for all four AAs.}
  \label{fig:PMF_flexiblebounds}
\end{figure}

For each of the four AA molecules, we performed 10 independent runs to evaluate equation~(\ref{eq:free_energy_difference}).  The averaged results for these approximate free energy barriers are shown in Figure~\ref{fig:PMF_flexiblebounds}. Even though the standard deviations are rather large, we see a clear separation between AA1 and AA2 on the one hand (average free energy difference \SI{11.6}{kJ/mol} and \SI{10.8}{kJ/mol}, respectively), and AA3 and AA4 on the other hand (average free energy difference \SI{3.5}{kJ/mol} and \SI{1.8}{kJ/mol}, respectively).
The number of samples (10) is too small to perform a stringent statistical analysis of this data.
However, together with the observations made from the force-distance profiles, we can still
draw the conclusion that AA1 and AA2 are better suited to prevent agglomeration than AA3 and AA4. This is in excellent agreement with the ranking that we obtain from the non-steered simulations. 

\subsection{Experimental results}
The results of the experimental rocking cell tests are shown in Table~\ref{tab:table_steered_runs}. To characterize the performance of each of the four inhibitor molecules, we indicate the MED that was required to prevent agglomeration. AA1 and AA2 are very close together, with an MED of 1.17\% and 1.00\%, respectively. It is questionable whether this small difference is meaningful, not least due to the discrete steps by which the concentration was increased during the testing procedure. We consequently consider these two molecules to perform equally well as anti-agglomerants. For AA3 and AA4, on the other hand, it was not possible to determine the MED, since they did not demonstrate inhibition up to highest concentration that was tested (3.00\%). We therefore consider them to perform badly as anti-agglomerants.

\begin{table}
 \begin{small}
 \begin{tabularx}{0.7\textwidth}{W W W W W}
  \toprule
  Concentration & AA1                     & AA2                     & AA3                   & AA4                   \\
  0.00\%        & \textcolor{red}{Fail}   & \textcolor{red}{Fail}   & \textcolor{red}{Fail} & \textcolor{red}{Fail} \\
  0.50\%        & --                      & \textcolor{red}{Fail}   & --                    & --                    \\
  0.67\%        & \textcolor{red}{Fail}   & --                      & --                    & --                    \\
  0.83\%        & \textcolor{red}{Fail}   & --                      & --                    & --                    \\
  1.00\%        & \textcolor{red}{Fail}   & \textcolor{green}{Pass} & \textcolor{red}{Fail} & \textcolor{red}{Fail} \\
  1.17\%        & \textcolor{green}{Pass} & --                      & --                    & --                    \\
  1.33\%        & \textcolor{green}{Pass} & --                      & --                    & --                    \\
  1.50\%        & --                      & \textcolor{green}{Pass} & \textcolor{red}{Fail} & --                    \\
  2.00\%        & --                      & \textcolor{green}{Pass} & \textcolor{red}{Fail} & \textcolor{red}{Fail} \\
  2.50\%        & --                      & \textcolor{green}{Pass} & \textcolor{red}{Fail} & --                    \\
  3.00\%        & --                      & --                      & \textcolor{red}{Fail} & \textcolor{red}{Fail} \\
  \bottomrule
 \end{tabularx}
 \end{small}
 \caption{Results of the experimental evaluation of the anti agglomeration performance. The AA concentration was increased until the system stopped to agglomerate, as described earlier. The concentration is indicated as volume percentage with respect to the volume of the water phase.}
 \label{tab:table_steered_runs}
\end{table}

Overall, the experimental results confirm the predictions made by the simulations: AA1 and AA2 are very close together and show good performance, whereas AA3 and AA4 are clearly worse. Even though the simulations are not yet capable of yielding quantitative predictions (i.e. to directly calculate the MED) and are limited to qualitative predictions, this does not belittle their utility. Relative comparisons among several molecules, in combination with a few quantitative experimental reference points, should allow to make reasonable quantitative predictions.

Once such a reference framework has been set up, it should then be possible to perform a systematic computational high-throughput screening of many molecules, exploiting scalable computational resources and going beyond the limitations and a purely lab-based approach.
On the one hand, this approach demonstrates the power of computational methods, and on the other hand underlines the necessity to complement them with experimental validation and calibration.

\section{Conclusions}

In this paper, we have used both computational and experimental methods to estimate the potency of four surfactant molecules to inhibit hydrate agglomeration. With respect to the simulations, we used both steered and non-steered MD to simulate the agglomeration process and the effect of the anti-agglomerants. With respect to the experiments, we used rocking cell measurements to determine the MED that prevents agglomeration. Based on these simulations and measurements, we could establish both a computational and an experimental ranking of the four molecules. We observed an excellent agreement between both rankings, indicating that simulations have become mature enough to accurately predict the performance of such molecules. Moreover, simulations on an atomistic level, as performed in this study, provide additional insights into the agglomeration process and the way in which the inhibitors prevent it that would not be accessible with purely experimental methods. For instance, we analyzed in this study the density profiles at the interface, the diffusion of the surfactants, and the orientation of their tails.

We do not aim at exploring why some of the tested surfactants work better than others, but limited us to a qualitative comparison of the performances.
Even though the simulations are not yet capable of yielding quantitative predictions, they represent a very powerful tool when combined with experimental work, as done in this study. The possibility to perform systematic computational high-throughput screenings of many molecules, exploiting scalable computational resources, allows to set up an efficient funnel approach where only the most promising candidates will eventually be synthesized and tested in the lab. This allows to go beyond a purely experimental approach where one has to synthesize and test every molecule in the laboratory, making research more efficient and scalable.

\begin{acknowledgement}
The authors thank Clariant for the financial support and to allow for the work to be published.
\end{acknowledgement}


\bibliography{references}

\end{document}